\documentclass[review,12pt]{elsarticle}
\usepackage{amssymb}
\usepackage{amsthm}
\usepackage{amsmath,amsfonts}
\usepackage{bm}

\usepackage{enumitem}
\usepackage{comment}
\usepackage{setspace}
\usepackage{gensymb}
\usepackage{upgreek}
\usepackage[colorlinks=true,
            linkcolor=red,
            urlcolor=red,
            citecolor=red]{hyperref}
\usepackage[nameinlink]{cleveref}
\crefname{figure}{Fig.}{Figs.}
\crefname{equation}{Eq.}{Eqs.}

\usepackage{svg}
\usepackage{flafter}
\usepackage{nomencl}
\usepackage{mdframed} % to create box around nomenclature
\makenomenclature

\usepackage{mathrsfs}
\usepackage{graphicx}
\usepackage{epstopdf}
\usepackage{float}
\usepackage{caption}
\usepackage{subcaption}
\usepackage{bm}
\usepackage{bbm}
\usepackage{soul}
\usepackage{accents}
\usepackage{xcolor}
\usepackage{courier} % To write in Courier font (to write code basically)
\usepackage{listings} % To include code
\usepackage{tabu} % To increase the vertical space between cells in tables
\usepackage{longtable}
\usepackage{changepage} % To use the adjustwidth feature that allows pushin wide tables to the left margin
\usepackage[margin=2cm]{geometry}% TO PLAY WITH THE MARGINS
\biboptions{sort&compress} %To make [1,2,3] be [1-3]
\journal{International Journal of Hydrogen Energy}

\makeatletter
\def\@author#1{\g@addto@macro\elsauthors{\normalsize%
    \def\baselinestretch{1}%
    \upshape\authorsep#1\unskip\textsuperscript{%
      \ifx\@fnmark\@empty\else\unskip\sep\@fnmark\let\sep=,\fi
      \ifx\@corref\@empty\else\unskip\sep\@corref\let\sep=,\fi
      }%
    \def\authorsep{\unskip,\space}%
    \global\let\@fnmark\@empty
    \global\let\@corref\@empty  %% Added
    \global\let\sep\@empty}%
    \@eadauthor={#1}
}
\makeatother

\begin{document}

\begin{frontmatter}

%% Title, authors and addresses

%% use the tnoteref command within \title for footnotes;
%% use the tnotetext command for theassociated footnote;
%% use the fnref command within \author or \address for footnotes;
%% use the fntext command for theassociated footnote;
%% use the corref command within \author for corresponding author footnotes;
%% use the cortext command for theassociated footnote;
%% use the ead command for the email address,
%% and the form \ead[url] for the home page:
%% \title{Title\tnoteref{label1}}
%% \tnotetext[label1]{}
%% \author{Name\corref{cor1}\fnref{label2}}
%% \ead{email address}
%% \ead[url]{home page}
%% \fntext[label2]{}
%% \cortext[cor1]{}
%% \address{Address\fnref{label3}}
%% \fntext[label3]{}

\title{Can dents and gouges compromise the structural integrity of hydrogen transport pipelines?}

%% use optional labels to link authors explicitly to addresses:
%% \author[label1,label2]{}
%% \address[label1]{}
%% \address[label2]{}

\author[Oxf]{Ratul Das}

\author[SHELL]{Bostjan Bezensek}

\author[Oxf]{Emilio Mart\'{\i}nez-Pa\~neda\corref{cor1}}
\ead{emilio.martinez-paneda@eng.ox.ac.uk}

\address[Oxf]{Department of Engineering Science, University of Oxford, Oxford OX1 3PJ, UK}

\address[SHELL]{Shell Research Ltd., 455 Union Street, Aberdeen, AB11 6DB, United Kingdom}

\cortext[cor1]{Corresponding author.}

\begin{abstract} % IJHE - 150 words
Repurposing natural gas pipelines for hydrogen transport requires understanding how external defects, like dents and gouges, affect structural integrity under H$_2$ exposure. To address this, we combine experiments with a new hydrogen embrittlement model aimed at large plastic straining scenarios, which integrates: (i) multi-trap hydrogen transport, (ii) finite-strain plasticity, and (iii) a hydrogen- and triaxiality-dependent damage law. Each constituent of the model is validated with experiments on X65 pipeline steel: (i) hydrogen permeation, (ii) full-scale pipe-indentation, and (iii) mechanical testing at different hydrogen and triaxiality levels. The validated model is used to study \textit{passive} (indent before H$_2$ exposure) and \textit{active} (indent with H$_2$) dents and gouges. Results reveal that hydrogen does not significantly increase the damage severity of those defects, unless hydrogen egress is completely precluded at the outer surface of a pipeline that is being pressurised internally and contains a pre-existing \textit{passive} dent with a gouge.\\ 
\end{abstract}

\begin{keyword}

Hydrogen embrittlement \sep dent \sep gouge \sep pipeline integrity \sep damage mechanics
%% keywords here, in the form: keyword \sep keyword

%% PACS codes here, in the form: \PACS code \sep code

%% MSC codes here, in the form: \MSC code \sep code
%% or \MSC[2008] code \sep code (2000 is the default)

\end{keyword}

\end{frontmatter}
% \setstretch{1}

% \linenumbers

%% main text
\section{Introduction}
\label{Introduction}

Hydrogen is vital to a successful global energy transition due to its potential to decarbonise hard-to-electrify sectors such as steelmaking, transportation, and power generation \cite{HOSSAINBHUIYAN20251026, ABA2024660, edwards2007hydrogen}. In the near future, steel pipelines are expected to play a vital role in establishing a transmission `hydrogen backbone', enabling efficient, large-scale distribution of high-pressure hydrogen gas (H$_2$) from production sites to end users. To this end, governments worldwide have favoured integrating hydrogen into the natural gas transport infrastructure, as it is 5 to 10 times cheaper to retrofit natural gas pipelines for H$_2$ transport than to build new, dedicated ones \cite{LIPIAINEN202331317,vreeburg2025potential,mielich2025europe}. However, the transport of hydrogen in new or repurposed steel pipelines is not without challenges \cite{LAUREYS2022104534,chalfoun2025tailored}.\\

Like most metallic materials, pipeline steels can undergo \emph{hydrogen embrittlement} - a notable reduction in ductility, fracture toughness, and fatigue crack growth resistance due to the absorption of hydrogen \cite{johnson1875ii,djukic2019synergistic,yu2024hydrogen,chen2025hydrogen}. Experimental studies have shown that lattice hydrogen concentrations below 1 part per million in weight (ppm wt) can reduce ductility (elongation at failure) by 50\% \cite{WANG2022144262, lee2011mechanical, bae2014effect, entezari2024experimental,Hoschke2023} and fracture resistance by 80\% \cite{san2012technical,briottet2018influence,BOUKORTT201819615}. Furthermore, existing natural gas pipelines have been installed over a period exceeding five decades, and therefore contain a wide range of defects, arising during manufacturing and welding or through decades of operation, often in aggressive environments. As shown by Mandal \textit{et al.} \cite{MANDAL2025923}, pre-existing defects play a fundamental role in governing the structural integrity of hydrogen transmission pipelines, as hydrogen accumulates in areas of high hydrostatic stresses and plastic strains, significantly reducing the local fracture resistance of the material.\\

Due to the vast combination of materials (including welds), conditions (H$_2$ purity and pressure) and pre-existing defects, assessing the structural integrity of natural gas pipelines exposed to hydrogen through experimentation is challenging. As a result, recent years have seen a growing interest in the application of numerical methods to map safe regimes of operation in hydrogen transmission pipelines, for the wide range of relevant conditions \cite{kristensen2020applications,xu2024engineering,QIN2025}. For example, Zhang and Tian \cite{zhang2022failure} combined finite element analysis and genetic algorithms to predict damage and bursting pressures of hydrogen pipelines. Mandal \textit{et al.} \cite{MANDAL2025923} combined thermo-mechanical weld process modelling and phase field-based hydrogen embrittlement simulations to predict the critical failure pressures of hydrogen transport pipelines. Wijnen \textit{et al.} \cite{WIJNEN2025113533} enriched this framework to provide a thermo-metallurgical description of welding, which enabled accounting for the microstructural heterogeneity of welds. The outcome of their coupled welding and phase field-based hydrogen embrittlement simulations was then integrated into \emph{Virtual} Failure Assessment Diagrams (FADs), which enable incorporating advanced modelling into simple engineering assessment protocols \cite{wijnen2025virtual}. Xu \textit{et al.} \cite{xu2025research} conducted coupled deformation-diffusion-fracture phase field calculations to understand the role that weld-related stress concentrators play in the structural integrity of hydrogen transmission pipelines. Ning \textit{et al.} \cite{ning2025finite} conducted coupled deformation-diffusion finite element calculations to understand the evolution of hydrogen content near weld defects under static and cyclic loading. Nazar and Proverbio \cite{nazar2025modeling} used phase field modelling to predict the interplay between hydrogen and fatigue damage in pipeline steels.  
However, these efforts have focused purely on assessing the role of internal defects, which are considered more critical as they are exposed to higher stresses and hydrogen concentrations. Similarly, experimental endeavours have also focused largely on replicating the conditions relevant to internal defects, typically by conducting tests on samples exposed to the targeted inner H$_2$ pipeline pressure \cite{10.1115/PVP2021-62045, 10.1115/PVP2011-57684, 10.1115/PVP2010-25825, osti_1884667}. However, external defects are common in natural gas pipelines, as they arise due to environmental and soil interactions (e.g., corrosion), as well as due to third-party damage - impact from agricultural and excavating equipment is one of the most frequent failure root causes in natural gas pipelines. As a result, the interplay between pipeline structural integrity and external defects such as dents and gouges has been studied extensively in natural gas pipelines, but not under the consideration of H$_2$. Only very recently, Zhao and Cheng \cite{ZHAO2024109651} have used a phase field model for hydrogen embrittlement to estimate under what conditions hydrogen-induced cracks would initiate in dented pipelines. However, their study was limited to shallow dent depths and assumed that hydrogen could not escape from the outer surface of the pipeline, an unlikely scenario given the size of the hydrogen atom. Therefore, there is a need to understand, under realistic conditions, how the interaction between hydrogen and external defects (dents, gouges) can compromise the integrity of H$_2$ transport pipelines, and to update integrity assessment methodologies accordingly. \\

In this work, we combine experiments and modelling to shed light on the synergistic effect of external defects and hydrogen. Both dents and gouges are considered, covering a wide range of indentation dent depths (up to 13.2\% of the outer diameter), as well as both active (indent in the presence of H$_2$) and passive (indent prior to H$_2$ exposure) denting scenarios. To simulate damage under conditions of large plastic straining, a new deformation-diffusion-damage model is presented, which combines large strain plasticity, hydrogen diffusion and trapping, and a triaxiality- and hydrogen-dependent damage evolution law. The model is validated with dedicated indentation experiments and against hydrogen permeation and mechanical tests from the literature. The validated model is employed to predict the failure of dented pipelines under a wide range of conditions, and answer the following pressing questions: (i) will dents and external gouges that are deemed safe in natural gas transmission cause catastrophic failure in hydrogen-carrying pipelines?, (ii) how does active indentation in a hydrogen-containing pipeline differ from passive indentation where the dent is generated prior to hydrogen uptake?, and (iii) what role does the outer condition of the pipeline (restricted vs unrestricted hydrogen egress) play?\\

The remainder of this paper is structured as follows. In Section \ref{Sec:Governing Equations}, the novel model for damage under hydrogen transport and large plastic straining is presented, including details of its numerical implementation. Then, in Section \ref{Sec:Experiments}, details of the experimental calibration and validation are provided. This spans the three constituents of the model: (i) the hydrogen diffusion and trapping characteristics, benchmarked against hydrogen permeation tests, (ii) the damage evolution law, defined based on experiments at different levels of hydrogen content and triaxiality, and (iii) the large-strain elastic-plastic formulation, validated against dedicated, full-scale denting experiments. Section \ref{Sec:Results} presents the results obtained for the analysis of a hydrogen transport pipeline containing dents and gouge defects. Both active and passive denting are considered, and the findings are discussed in the context of the literature and in terms of their implications for safe hydrogen transport. Finally, Section \ref{Sec:Conclusions} summarises the manuscript and provides concluding remarks.

\section{A damage mechanics model for hydrogen-assisted damage in elastic-plastic solids undergoing large deformations}
\label{Sec:Governing Equations}

We proceed to describe the newly developed model to predict hydrogen-assisted damage under large plastic straining. First, in Section \ref{Sec:Htransport}, the hydrogen transport formulation is described, with a particular emphasis on the ability to capture the role of multiple trap types, and the interplay between plasticity (dislocations) and hydrogen trapping, which is significant for dents. Then, the constitutive model adopted to capture large plastic straining is introduced in Section \ref{sec:GradientPlasticity}. Finally, in Section \ref{Sec:DamageLaw}, a triaxiality- and hydrogen-dependent damage law is presented.

\subsection{Hydrogen diffusion and multi-trapping}
\label{Sec:Htransport}

The relevant theory of hydrogen diffusion in steel is briefly reviewed in this section, including four salient features critical to the problem under consideration: (i) a hydrostatic stress-gradient-dependent flux resulting from the reduction in the chemical potential of lattice sites under tensile stress \cite{osti_4288898,osti_4121406,SOFRONIS1989317}, (ii) the trapping of hydrogen, considering multiple trap types (grain boundaries, dislocations, carbides) \cite{DADFARNIA201110141}, (iii) mechanistic and phenomenological relationships between plasticity and hydrogen trapped at dislocations \cite{Kumnick198033,LIN2024146175,SOFRONIS2001857,FERNANDEZSOUSA2020253}, and (iv) the depletion of lattice sites because of fast creation of traps during plastic deformation \cite{DIAZ2025111007,KROM1999971}.\\

\begin{figure}[]
\centering
\includegraphics[width=6 in]{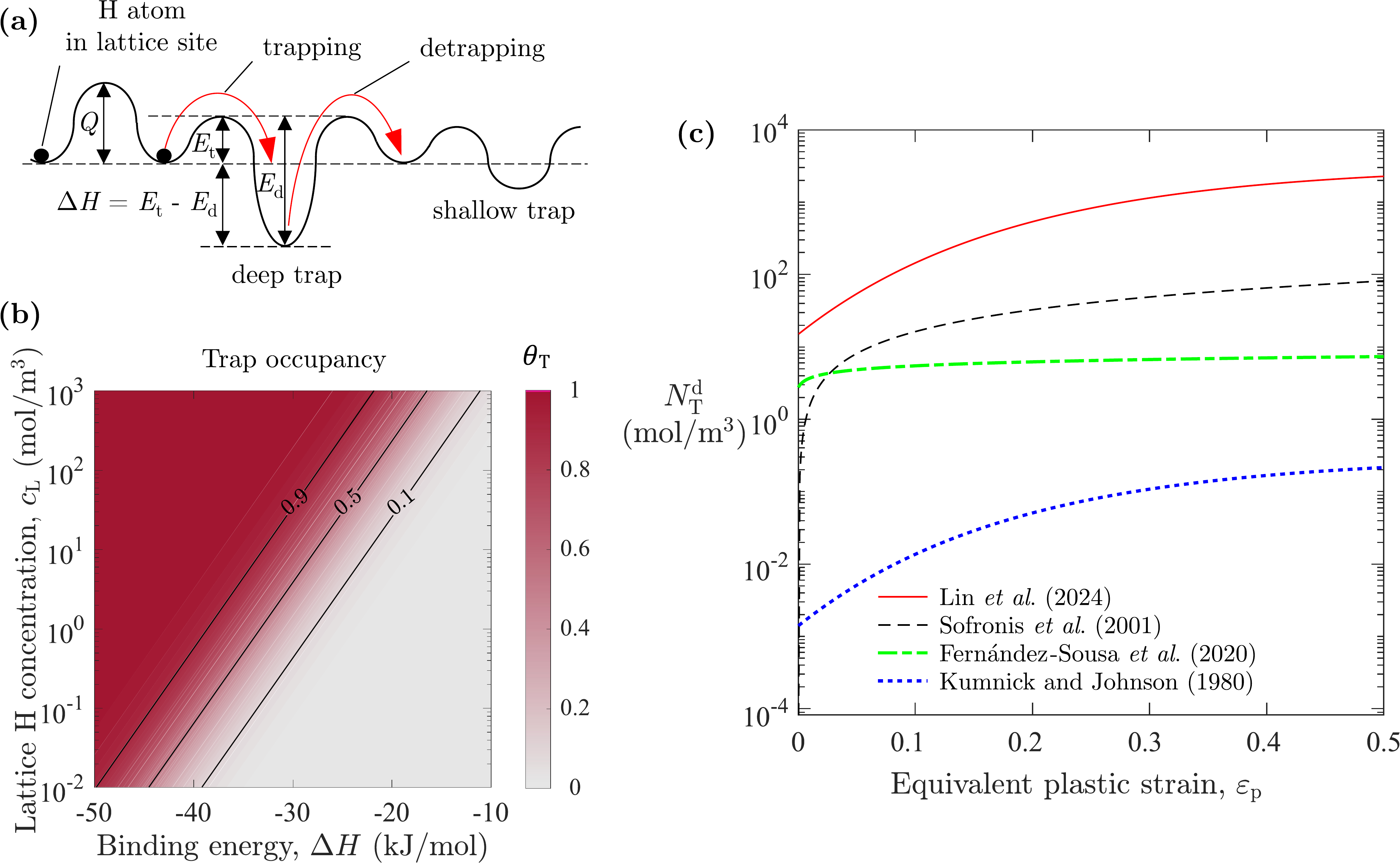}
\caption{Hydrogen diffusion and trapping; (a) Schematic of the energy landscape as a function of the lattice activation energy $Q$, the trapping ($E_\text{t}$) and detrapping ($E_\text{d}$) energies, and the trap binding energy $\Delta H$ \cite{garcia2024tds}; (b) trap occupancy ($\theta_{\text{T}}$) as a function of lattice hydrogen concentration $c_\text{L}$ and trap binding energy $\Delta H$, as dictated by Oriani's equilibrium \cite{ORIANI1970147}; (c) evolution of the dislocation trap density $N^{\text{d}}_{\text{T}}$ as a function of equivalent plastic strain $\varepsilon_\mathrm{p}$, as predicted with four representative phenomenological \cite{Kumnick198033,LIN2024146175} and mechanistic \cite{SOFRONIS2001857,FERNANDEZSOUSA2020253} models.}
\label{F_Energy_Landscape}
\end{figure}

Hydrogen ingress into a metal exposed to H$_2$ is the result of molecular dissociation, adsorption and absorption. The absorbed hydrogen atoms can diffuse through the lattice or be `trapped' at defects \cite{cupertino2023hydrogen}, with the hydrogen binding energy of each defect ($\Delta H$) governing trapping characteristics. Defects with a high absolute value of $\Delta H$ are known as `deep traps', while low $|\Delta H|$ values describe shallow traps, as illustrated in Fig. \ref{F_Energy_Landscape}(a). Considering the general case of an alloy with multiple trap defects (carbides, dislocations, grain boundaries, \textit{etc.}), the total hydrogen concentration $c$ is additively decomposed into the hydrogen concentration in the lattice ($c_\text{L}$) and the hydrogen concentration in trapping sites ($c_\text{T}^\text{i}$), such that $c = c_\text{L}+\sum\nolimits_\text{i} c_\text{T}^\text{i}$. Here, the superscript `i' is used to denote the trap type (e.g., `i = d' for dislocations, `i = gb' for grain boundaries, `i = c' for carbides).  Considering the role that hydrostatic stresses play in increasing atomic spacing in the lattice, mass balance dictates that the transport of hydrogen shall be described by \cite{SOFRONIS1989317},
\begin{equation}
\frac{\partial c_{\text{L}}}{\partial t} + \sum\nolimits_\text{i}\frac{\partial c_\mathrm{T}^\text{i}}{\partial t} + \nabla \cdot \left( -D_{\text{L}} \nabla  c_{\text{L}} +  \frac{D_{\text{L}} V_{\text{H}}}{RT}  c_{\text{L}} \nabla \sigma_{\text{h}} \right) = 0 \, .
\label{Eq:modFick}
\end{equation}

\noindent Here, $D_\text{L}$ is the lattice diffusion coefficient, $V_{\text{H}}$ denotes the partial molar volume of hydrogen in solid solution, $R$ is the universal gas constant, $T$ is the absolute temperature, and $\sigma_{\text{h}} = \text{tr}(\boldsymbol{\sigma})/3$ denotes the hydrostatic stress, with $\bm{\sigma}$ being the Cauchy stress tensor. The solution variable in Eq. (\ref{Eq:modFick}) is $c_\text{L}$ and thus, a relationship must be established between the trapped and lattice hydrogen concentration. Denoting $N_{\text{L}}$ and $N_{\text{T}}^\text{i}$ as the densities of lattice and trap sites, respectively, and assuming Oriani's equilibrium \cite{ORIANI1970147}, the occupancy of lattice ($\theta_{\text{L}} = c_{\text{L}}/N_{\text{L}}$) and trapping ($\theta_\text{T}^\text{i} = c_{\text{T}}^\text{i}/N_{\text{T}}^\text{i}$) sites can be related as,
\begin{equation}
    \frac{\theta_\text{T}^\text{i}}{1-\theta_\text{T}^\text{i}} = \frac{\theta_{\text{L}}}{1-\theta_{\text{L}}} K^\text{i} = \frac{\theta_{\text{L}}}{1-\theta_{\text{L}}} \exp \left(\frac{-\Delta H^\text{i}}{R T}\right) \, ,
    \label{Eq:Oriani}
\end{equation}

\noindent where $K^\text{i}$ is the equilibrium constant. The implications of Oriani's equilibrium are shown in Fig. \ref{F_Energy_Landscape}(b), by plotting the trap occupancy ($\theta_\text{T}$) for a wide range of trap binding energies ($\Delta H$) and lattice hydrogen concentrations ($c_\text{L}$), upon considering Eq. (\ref{Eq:Oriani}) and $c_\text{L}= \theta_{\text{L}} N_{\text{L}}$, with the lattice site density being $N_{\text{L}} = \beta {N_{\text{A}}}/{V_{\text{M}}} = 5.1 \times 10^{29} \, \text{sites/m}^3$ (or $8.47 \times 10^{5} \, \text{mol}/\text{m}^3$) for bcc materials, where the molar volume $V_{\text{M}} = 7.116 \times 10^{-6} \, \text{m}^3/\text{mol}$, $N_\text{A}$ is Avogadro's constant, and $\beta$ is the number of interstitial sites per solvent atom (6 in a bcc lattice). As shown in Fig. \ref{F_Energy_Landscape}(b), deep traps at $\Delta H^\text{i} \leq - 50$ kJ/mol saturate at a very low lattice hydrogen concentration ($c_\text{L} \leq 0.02 $ mol/m$^3$) whereas shallow traps with $\Delta H^\text{i} \geq - 20$ kJ/mol are effectively empty unless $c_\text{L} \geq 10 $ mol/m$^3$. The trap density $N_\text{T}$ is assumed to be constant for most trap types but evolves for dislocations, as a result of dislocation density evolution with plastic straining. Hence, considering the role of the plastic strain rate in creating dislocation trap sites and noting that $\theta_\text{L} << 1$ in bcc steels \cite{DIAZ2025111007}:
\begin{equation}
\sum\nolimits_\text{i}\frac{\partial c_\mathrm{T}^\text{i}}{\partial t} = % limits of summation in inline math mode
\sum\nolimits_i \left[ \frac{K^{\text{i}} N_{\mathrm{T}}^{\text{i}}/N_{\mathrm{L}}}{(1+K^{\text{i}}c_{\mathrm{L}}/N_{\text{L}})^2}\frac{\partial c_{\text{L}}}{\partial t} + \theta_{\mathrm{T}}^{\text{i}}\frac{\text{d} N_{\text{T}}^{\text{i}}}{\text{d} \varepsilon_\mathrm{p}}\frac{\text{d} \varepsilon_\mathrm{p}}{\text{d} t} \right].
\label{Eq:HTransportmult}
\end{equation}

\noindent Eqs. (\ref{Eq:modFick}) and (\ref{Eq:HTransportmult}) together fully describe the diffusion of hydrogen in steel, accounting for the role of hydrostatic stresses, trapping at different defect types, and the creation of trap types with increasing dislocation density (plastic deformation). The interplay between plastic deformation and trapping is key in the analysis of denting and other processes involving high strain levels. Different models have been proposed to describe the evolution of $N_\text{T}^\text{d}$ with equivalent plastic strain $\varepsilon_\text{p}$, phenomenological \cite{Kumnick198033,LIN2024146175} or mechanistic \cite{SOFRONIS2001857,FERNANDEZSOUSA2020253}. The latter define the trap density as a function of the dislocation density: $N_\text{T}^\text{d} = \sqrt{2}\rho/a$, with the lattice parameter being $a = 0.2867$ nm for bcc materials, and provide a physically-based relationship to link dislocation density $\rho$ with equivalent plastic strain $\varepsilon_\text{p}$. For example, Sofronis \textit{et al.} \cite{SOFRONIS2001857} assumed that $\rho$ varies linearly with $\varepsilon_\text{p}$ until a saturation value: 
\begin{equation}
\rho = \begin{cases} 
\rho_0 + \gamma \varepsilon_\text{p} & \text{for } \varepsilon_\text{p} < 0.5, \\
10^{16} & \text{for } \varepsilon_\text{p} \geq 0.5,
\end{cases}
\end{equation}
\noindent where $\rho_0=10^{10}$ line length/m$^3$ corresponds to the dislocation density for the annealed material and the fitting constant $\gamma = 2 \times 10^{16}$ line length/m$^3$. On the other hand, Fern\'andez-Sousa \textit{et al.} \cite{FERNANDEZSOUSA2020253} used Taylor's dislocation model, which, in the absence of gradient effects/GNDs, results in,
\begin{equation}
    \rho= \left( \frac{\sigma_\text{y0} f( \varepsilon_\text{p}) }{0.5M \mu b} \right)^2 \, ,
\end{equation}

\noindent where $M$ is the Taylor factor (equal to 2.9 for bcc metals), $\mu$ is the shear modulus, $b$ is Burgers vector (0.248 nm for bcc metals), $\sigma_\text{y0}$ is the initial yield stress and $f(\varepsilon_\text{p})$ is the hardening law, which in this paper is taken as $f(\varepsilon_\text{p})=(1+E\varepsilon_\text{p}/\sigma_\text{y0})^N$ with $N$ being the strain hardening exponent ($0\leq N \leq 1)$ and $E$ denoting Young's modulus. From a phenomenological perspective, Kumnick and Johnson \cite{Kumnick198033} were the first to conduct permeation experiments with samples deformed to different levels. Their data in pure iron can be approximated with a relationship of the type,
\begin{equation}
\log {N}_{\text{T}}^{\text{d}}=A-2.33 \exp \left(-5.5 \varepsilon_{\text{p}}\right) \, ,
\label{Eq:plasticstrainevo_gen}
\end{equation}

\noindent with $A=23.26$ providing the best fit to the data \cite{SOFRONIS1989317}. Recently Lin \textit{et al.} \cite{LIN2024146175} have conducted similar experiments on X65 pipeline steel and found $A$ to be 27.29. The evolution of $N_\text{T}^\text{d}$ with equivalent plastic strain $\varepsilon_\text{p}$ obtained from these four approaches is shown in Fig. \ref{F_Energy_Landscape}(c). It can be seen that the highest $N_\text{T}^\text{d}$ values are obtained with the expression derived from the study of Lin \textit{et al.} \cite{LIN2024146175}. Since this will result in a higher hydrogen content and thus higher potential susceptibility to hydrogen embrittlement, this is the dislocation trap density evolution considered in the calculations of this study, taking a conservative approach.

\subsection{Finite strain isotropic plasticity}
\label{sec:GradientPlasticity}

A suitable description for large strains is adopted by using the deformation gradient,
\begin{equation}
    \bm{F} = \bm{I} + \nabla \mathbf{u}  \, , 
\end{equation}
\noindent where $\bm{I}$ is the identity matrix and $\mathbf{u}$ the displacement vector, the solution variable for the mechanical problem. As is common in finite strain plasticity \cite{SIMO1985221}, the additive multiplicative decomposition of the deformation gradient into its elastic and plastic parts is adopted,
\begin{equation}
    \bm{F} = \bm{F}^{\mathrm{e}} \, \bm{F}^{\mathrm{p}} \, ,
\end{equation}

\noindent with $\bm{F}^{\mathrm{p}}$ being obtained through the flow rule, which will be of the form
\begin{equation}
\dot{\bm{F}}^{\mathrm p} = \bm{L}^{\mathrm{p}} \bm{F}^{\mathrm{p}} \, , \,\,\,\,\,\,\,\, \bm{F}^{\mathrm{p}} (t=0) = \bm{I} \, ,
\end{equation}

\noindent where $\bm{L}^{\mathrm{p}}$ is the plastic velocity gradient. In associative $J_2$ isotropic plasticity, where the plastic spin is neglected, one can formulate $\bm{L}^{\mathrm{p}}$ as a function of a plastic multiplier $\lambda$ and a flow direction $\bm{N}$, such that $\bm{L}^{\mathrm{p}}=\dot{\lambda} \bm{N}$, and consequently, the equivalent plastic strain becomes
\begin{equation}
   \dot{\varepsilon}_\text{p} = \sqrt{\frac{2}{3} \bm{L}^{\mathrm{p}} : \bm{L}^{\mathrm{p}}} \, , \,\,\,\,\,\,\, \varepsilon_\text{p} (t) = \int_0^t  \sqrt{\frac{2}{3} \bm{L}^{\mathrm{p}} : \bm{L}^{\mathrm{p}}} \, \text{d} t \, .
\end{equation}

\noindent The flow direction $\bm{N}$ is defined as 
\begin{equation}
    \bm{N} = \frac{3}{2} \frac{\bm{\tau'}}{\sqrt{\bm{\tau'}:\bm{\tau'}}} \, ,
\end{equation}

\noindent with $\bm{\tau'}$ being the deviatoric part of the Kirchhoff stress tensor. An exponential mapping algorithm is used to solve the plastic flow rule, considering the following yield condition
\begin{equation}
    f \left( \bm{\tau}, \sigma_{\text{y}} \right) = \sqrt{\frac{2}{3} \bm{\tau}' : \bm{\tau}'} - \sigma_{\text{y}} \left( \varepsilon_\text{p} \right) = 0 .
\end{equation}
\noindent where $\sigma_{\text{y}}$ is the current yield stress, which evolves in agreement with the following isotropic hardening law
\begin{equation}
\sigma_{\text{y}} \left( \varepsilon_\text{p} \right)=\sigma_{\text{y0}}\left(1+\frac{E\varepsilon_\mathrm{p}}{\sigma_\text{y0}}\right)^N,
\label{Eq_SwiftLaw}
\end{equation}
where $\sigma_{\text{y0}}$ is the initial yield strength and $N$ is the strain hardening exponent. Relevant Cauchy stress quantities can be obtained by noting that $\bm{\tau}=J \bm{\sigma}=\text{det} (\bm{F}) \bm{\sigma}$, where $\bm{\sigma}$ is the Cauchy stress tensor. Accordingly, a von Mises effective stress can be defined as $\sigma_\text{e}=\sqrt{(3/2) \bm{\sigma}' : \bm{\sigma}'}$, and the balance equation can be expressed as,
\begin{equation}
   \nabla \cdot \left[ \left(1-D \right) \bm{\sigma} \right] = \mathbf{0} \, ,
   \label{Eq:MomentumBalance}
\end{equation}

\noindent with $D$ being a damage variable ($0\leq D \leq 1)$, whose evolution is defined below. \\

\subsection{A triaxiality- and hydrogen-dependent damage law}
\label{Sec:DamageLaw}

The failure of metals subjected to high levels of deformation and hydrogen is predicted using a continuum damage mechanics formulation, with a scalar internal damage variable $D$ introduced to capture the degree of material degradation ($0\leq D \leq 1)$. Intact material points are described by $D=0$, while $D=1$ describes fully cracked material points, with intermediate values providing an estimate of the degree of microcracking that has occurred (i.e., $D=0.5$ signifying that 50\% of the load-bearing area of a microstructural representative volume element contains cracks). Damage irreversibility is enforced such that $\dot{D} \ge 0$. The approach is local and follows the work by Kim and co-workers \cite{oh2007development, SEO2022107371}. While local continuum damage mechanics models can lead to mesh-dependent localisation, this is not the case for the low degrees of damage attained in this study (see Section \ref{Sec:Results}). The damage variable $D$ incorporates the influence of both hydrogen and triaxiality on fracture. The latter is often neglected in hydrogen-assisted cracking studies where fracture is brittle or quasi-brittle, and plasticity is localised to the crack tip. However, in this case, failure would occur in the presence of very significant global plastic deformation, and thus triaxiality effects might be present. Defining triaxiality as the ratio between the hydrostatic stress and the effective von Mises stress ($\eta=\sigma_\mathrm{h}/\sigma_\mathrm{e}$), the damage variable is estimated as,
\begin{equation}
D = \int_{0}^{{\varepsilon}_{\mathrm{p}}} 
\frac{\mathrm{d}{\varepsilon}_{\mathrm{p}}}{f (c) g (\eta) \,\varepsilon_{\mathrm{f}}^0} \, ,
\label{Eq:DamageEvolution}
\end{equation}

\noindent where $\varepsilon_\text{f}^0$ is the failure strain in the absence of triaxiality and hydrogen effects, $g (\eta )$ is the triaxiality function that incorporates the drop in failure strain with increasing triaxiality levels, and $f(c)$ is the hydrogen degradation function. Accordingly, the degraded failure strain is $\varepsilon_\text{f}=f(c) g (\eta) \varepsilon_\text{f}^0$. To be on the conservative side, we have chosen to define $f(c)$ as a function of the \emph{total} hydrogen concentration. While some studies have shown that the diffusible hydrogen governs hydrogen-assisted failures \cite{guedes2020role}, others have shown that the irreversibly trapped hydrogen can also lead to embrittlement \cite{harris2018elucidating}. The lattice hydrogen concentration is expected to be low near the outer surface of the dent, as it is generally free to exit the pipeline, but the hydrogen trapped in dislocations will be significant and a large fraction of the total hydrogen content (particularly in bcc alloys due to their low lattice solubility). Hence, by considering the total hydrogen content, the most detrimental scenario is simulated. Following the work by Mandal \textit{et al.} \cite{MANDAL2025923}, an exponential form is adopted for the hydrogen degradation function:
\begin{equation}
f(c) = \chi \exp(\beta c) + \xi  \, ,
\label{Eq: fcgeneral}
\end{equation}
\noindent with $\chi$, $\beta$ and $\xi$ being material parameters, to be calibrated against experiments (see Section \ref{Sec:Experiments}). The triaxiality dependency is captured by means of the following triaxiality function \cite{oh2007development},
\begin{equation}
g \left(\eta  \right)  = q_1 \exp \left(q_2 \eta \right) + q_3 \, ,
\label{Eq: efgeneral}
\end{equation}

\noindent with $\eta=\sigma_\text{h}/\sigma_\text{e}$ and the coefficients $q_1$, $q_2$, and $q_3$ defined by benchmarking against experiments conducted on X65 pipeline steel at different triaxiality conditions.\\

It should be acknowledged that the damage model adopted here is a phenomenological one, which does not aim at resolving the mechanisms underlying hydrogen-assisted cracking; e.g., unlike Refs. \cite{serebrinsky2004quantum,martinez2016strain,nagao2018hydrogen,shishvan2020hydrogen,cupertino2024suitability}. As such, the model needs experimental calibration under relevant conditions, as conducted in Section \ref{Sec:DamageLawValidation}.

\section{Experimental benchmarking} 
\label{Sec:Experiments}

The coupled deformation-diffusion-fracture model presented in Section \ref{Sec:Governing Equations} is now particularised to the analysis of pipeline steels through a multi-step experimental benchmarking process. First, the hydrogen diffusion and trapping characteristics are quantitatively correlated with electro-permeation experiments in X65 pipeline steel (Section \ref{sec:HdiffusionExpt}). Subsequently, the hydrogen- and triaxiality-dependent damage law is calibrated against mechanical tests on pipeline steel samples exposed to different triaxiality and hydrogen environment scenarios (Section \ref{Sec:DamageLawValidation}). Finally, in Section \ref{Sec:Ind_H_free}, we report on full-scale indentation experiments on a hydrogen-free X65 pipeline, which serve as a robust validation of the constitutive model. This step-by-step validation of each of the features of the modelling framework provides a sound foundation for the coupled simulations that follow in Section \ref{Sec:Results}, aimed at shedding light on the interplay between hydrogen, pipeline integrity and external defects such as dents and gouges. 

\subsection{Validation of hydrogen diffusion and trapping predictions}
\label{sec:HdiffusionExpt} 

\begin{figure}[tb]
\centering
\includegraphics[width=5 in]{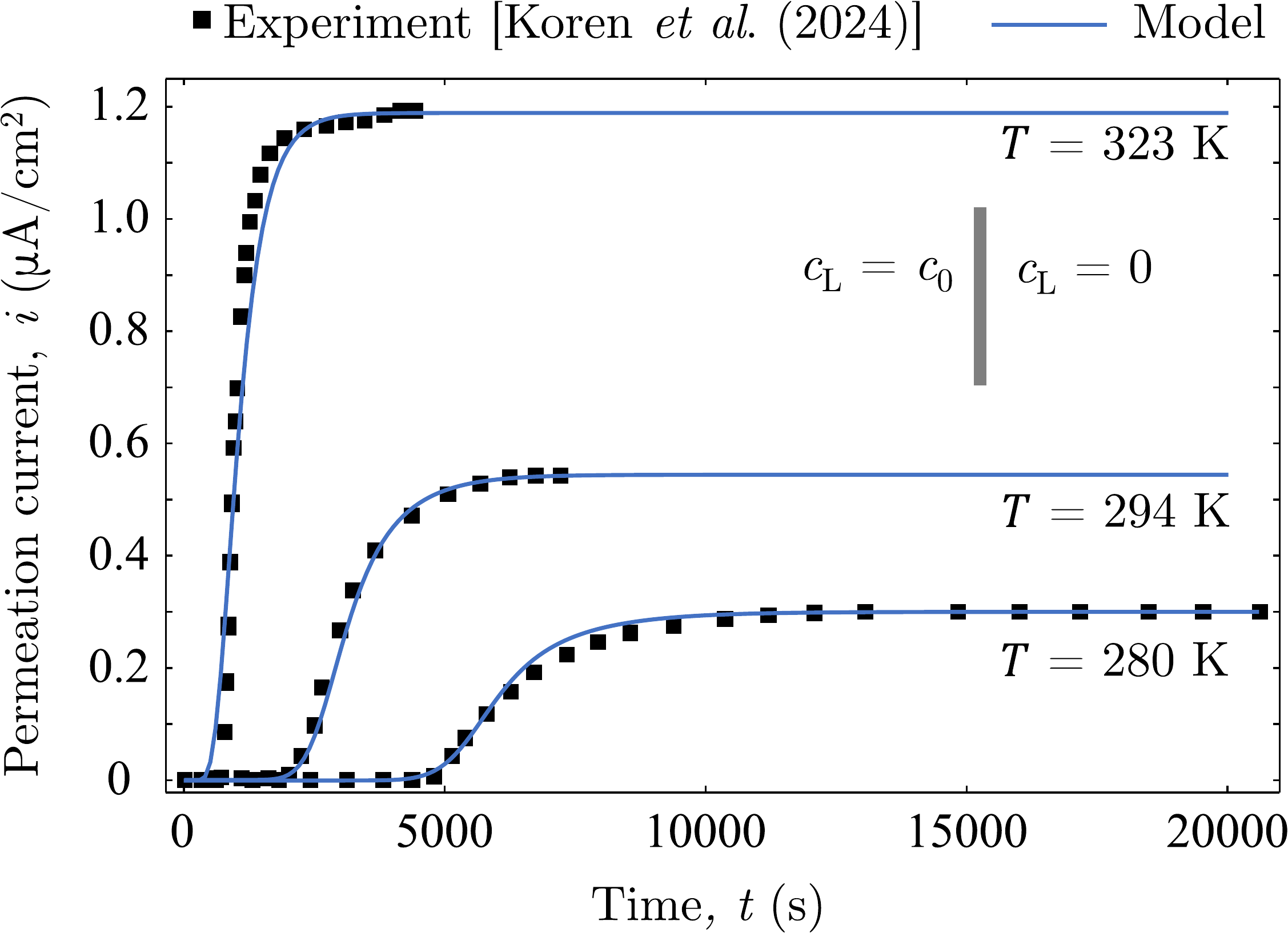}
\caption{Comparison of the simulated and experimentally measured \cite{KOREN20241157} permeation currents in X65 pipeline steel at various temperatures. A schematic of the numerical boundary value problem mimicking the conditions of electro-permeation tests is included in the inset.}
\label{F_EP_Calibration}
\end{figure}

The hydrogen transport model is particularised to X65 pipeline steel by benchmarking against the electro-permeation experiments conducted by Koren \textit{et al.} \cite{KOREN20241157} at three different temperatures (280, 294 and 323 K). This will enable establishing appropriate diffusion and trapping characteristics. Lattice diffusivity parameters are those of bcc iron \cite{KIUCHI198629}; i.e., $D_{\text{L}} = 7.23 \times 10^{-8} \exp(-Q/RT) \, \text{m}^2/\text{s}$ with $Q = 5690 \, \text{J/mol}$. The initial dislocation trap density is taken to be $N_{\text{T}}^{\text{d0}} = 15.1 \, \text{mol}/\text{m}^3$, as per the experiments by Lin \textit{et al.} \cite{LIN2024146175}. To establish the remaining trapping parameters, a numerical model of the permeation setup is developed (see the inset of Fig. \ref{F_EP_Calibration}). In such a model, a constant surface hydrogen concentration is prescribed ($c_\text{L} = c_0$) on the entry side, with its value given by,
\begin{equation}
c_0 = \frac{L}{F D_{\text{L}}} \, i_{\text{ss}} \, ,
\end{equation}

\noindent where $F$ is Faraday's constant, $L$ is the sample (membrane) thickness, and $i_\text{ss}$ is the experimentally measured steady state current density. On the exit side, a Dirichlet boundary condition $c_\text{L}=0$ is imposed, mimicking the experimental setup, and the permeating hydrogen flux $j$ is measured and reported as a transient current density: $i(t) = Fj(t)$. The results obtained are shown in Fig. \ref{F_EP_Calibration}. As it can be observed, an excellent agreement with experiments is attained (across three different temperatures) for the trap binding energies and densities reported in Table \ref{Tab:X65HTP}. As is common with pipeline steels, three trap types are assumed: dislocations, carbides and grain boundaries.

\begin{table}[H]
\centering
\caption{Experimentally-calibrated trap densities ($N_\text{T}^\text{i}$) and binding energies ($\Delta H^\text{i}$) for the three trap types considered: carbides ($N_\text{T}^\text{c}$, $\Delta H^\text{c}$), dislocations ($N_\text{T}^\text{d}$, $\Delta H^\text{d}$), and grain boundaries ($N_\text{T}^\text{gb}$, $\Delta H^\text{gb}$). For the case of dislocation traps, $N_\text{T}^\text{d0}$ denotes the initial trap density (in the unloaded configuration), with the trap density evolving with plastic strain as per Eq. (\ref{Eq:plasticstrainevo_gen}).}

\begin{tabular}{|>{\centering\arraybackslash}p{1.5cm}|
                >{\centering\arraybackslash}p{1.5cm}|
                >{\centering\arraybackslash}p{1.5cm}|
                >{\centering\arraybackslash}p{1.5cm}|
                >{\centering\arraybackslash}p{1.5cm}|
                >{\centering\arraybackslash}p{1.5cm}|}
\hline
\multicolumn{3}{|c|}{Trap densities [mol/m$^3$]} &
\multicolumn{3}{c|}{Trap binding energies [kJ/mol]} \\
\hline
$N_{\text{T}}^{\text{c}}$ & $N_{\text{T}}^{\text{d0}}$ & $N_{\text{T}}^{\text{gb}}$ &
$\Delta H^\text{c}$ & $\Delta H^\text{d}$ & $\Delta H^\text{gb}$ \\
\hline
2100 & 15.1 & 0.19 & $-20$ & $-28$ & $-50$ \\
\hline
\end{tabular}

\label{Tab:X65HTP}
\end{table}

\subsection{Damage sensitivity to hydrogen and triaxiality}
\label{Sec:DamageLawValidation}

\begin{figure}[tb]
\centering
\includegraphics[width=6.45in]{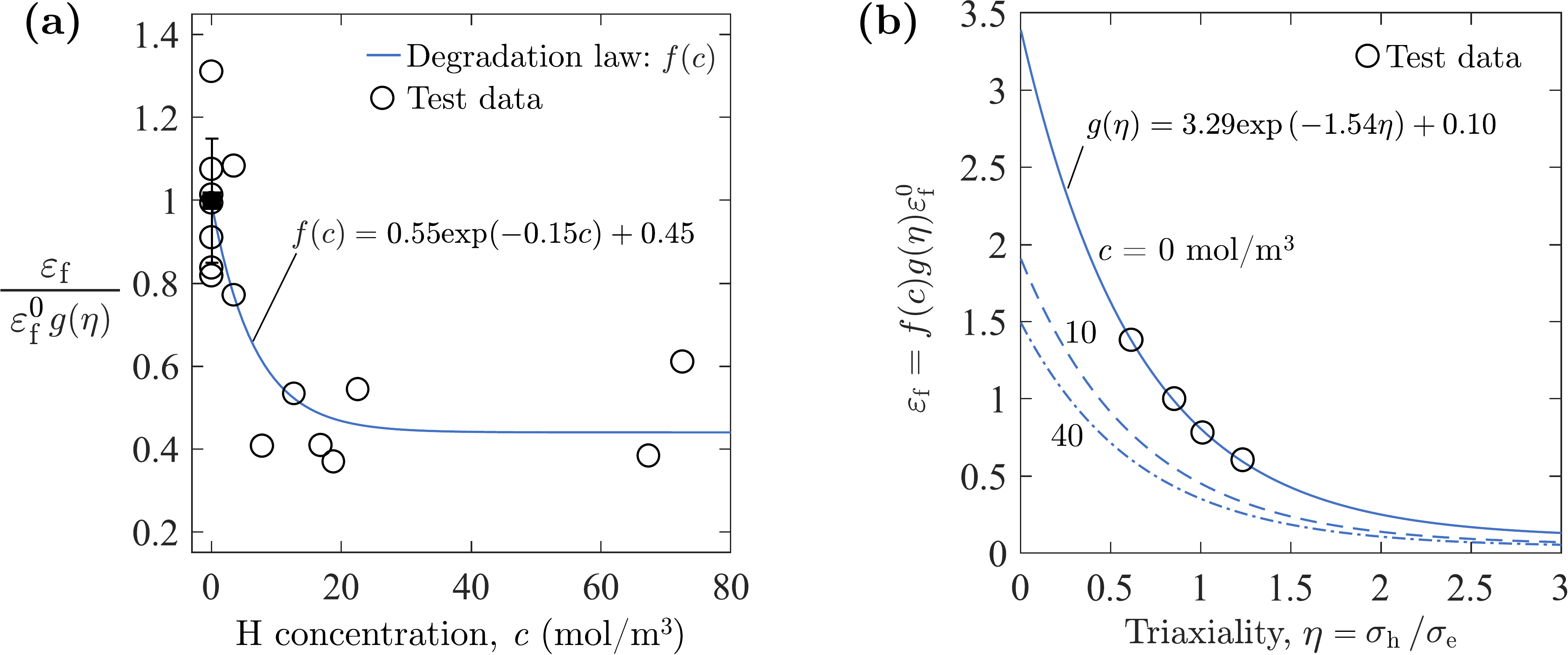}
\caption{Calibration of the hydrogen- and triaxiality-dependent damage law: (a) definition of the hydrogen degradation function $f(c)$ against experiments on pipeline steels \cite{WANG2022144262, ZHANG201729888, ZHOU201922547, HAN201922380, DONG20099879}, showing the failure strain normalised with the hydrogen-free one ($\varepsilon_\text{f}^0$) as a function of total hydrogen content; and (b) definition of the triaxiality function against experiments on notched hydrogen-free pipeline steel samples \cite{oh2007development}, showing the failure strain for different triaxiality levels ($\eta=\sigma_\text{h}/\sigma_\text{e}$) and including for completeness prediction curves including the effect of hydrogen too. In subfigure (a), the black square shows the average fracture strain in air, with the error bar indicating the standard deviation.}
\label{F_Degradation_Law}
\end{figure}

We proceed to particularise the hydrogen degradation function, Eq. (\ref{Eq: fcgeneral}), and the triaxiality-dependence of the fracture strain, Eq. (\ref{Eq: efgeneral}), for the case of pipeline steels. As shown in Fig. \ref{F_Degradation_Law}(a), the former is defined by fitting the strain at failure reported in uniaxial stress-strain tests from a range of literature works on pipeline steels \cite{WANG2022144262, ZHANG201729888, ZHOU201922547, HAN201922380, DONG20099879}. We emphasise that, to be on the conservative side, the hydrogen degradation law $f(c)$ is here calibrated against the total hydrogen concentration $c$. A good agreement is attained with the following hydrogen degradation law,
\begin{equation}
f(c) = 0.55 \exp(-0.15 ~c) + 0.45 \, ,
\label{Eq: H degradation}
\end{equation}

\noindent that is, with $\chi=0.55$, $\beta=-0.15$, and $\xi=0.45$.\\

The triaxiality dependence is incorporated in the damage formulation through Eqs. (\ref{Eq:DamageEvolution}) and (\ref{Eq: efgeneral}). Triaxiality effects on metallic fracture are intrinsically linked to void growth mechanisms that are typically not relevant to the brittle fracture of hydrogen-embrittled steels. However, these might be relevant in the damage arising due to large-scale denting, where hydrogen contents are low, and there is significant plastic straining. Thus, consistent with our approach of providing conservative estimations, triaxiality effects are accounted for. To this end, the experiments by Oh \textit{et al.} \cite{oh2007development} on hydrogen-free notched X65 pipeline steel samples are used as a benchmark (see Fig. \ref{F_Degradation_Law}(b)), providing the following expression,
\begin{equation}
g \left(\eta  \right)  = 3.29 \exp \left(-1.54 \eta \right) + 0.1 \, .
\end{equation}

\noindent Hence, $q_1=3.29$, $q_2=-1.54$ and $q_3=0.1$. In addition to the fit to experiments on notched samples exhibiting various triaxiality levels, Fig. \ref{F_Degradation_Law}(b) also includes two curves showing the combined degradation due to hydrogen and triaxiality, showing notable sensitivity to changes in hydrogen content and triaxiality levels, particularly for low values of $c$ and $\eta=\sigma_\text{h}/\sigma_\text{e}$.

\subsection{Full-scale pipeline denting: experiments and simulations}
\label{Sec:Ind_H_free}

The last stage of the validation process involves conducting dedicated denting experiments in an X65 pipeline and benchmarking the resulting dent profile with the predictions of the numerical model. We start by briefly describing the experiment. Then, we summarise details of the numerical model developed, which will be the one used to assess the interplay between hydrogen and external defects (dents, gouges). Finally, experimental measurements and numerical predictions are compared. 

\subsubsection{Experimental testing}

\begin{figure}[]
\centering
\includegraphics[width=6 in]{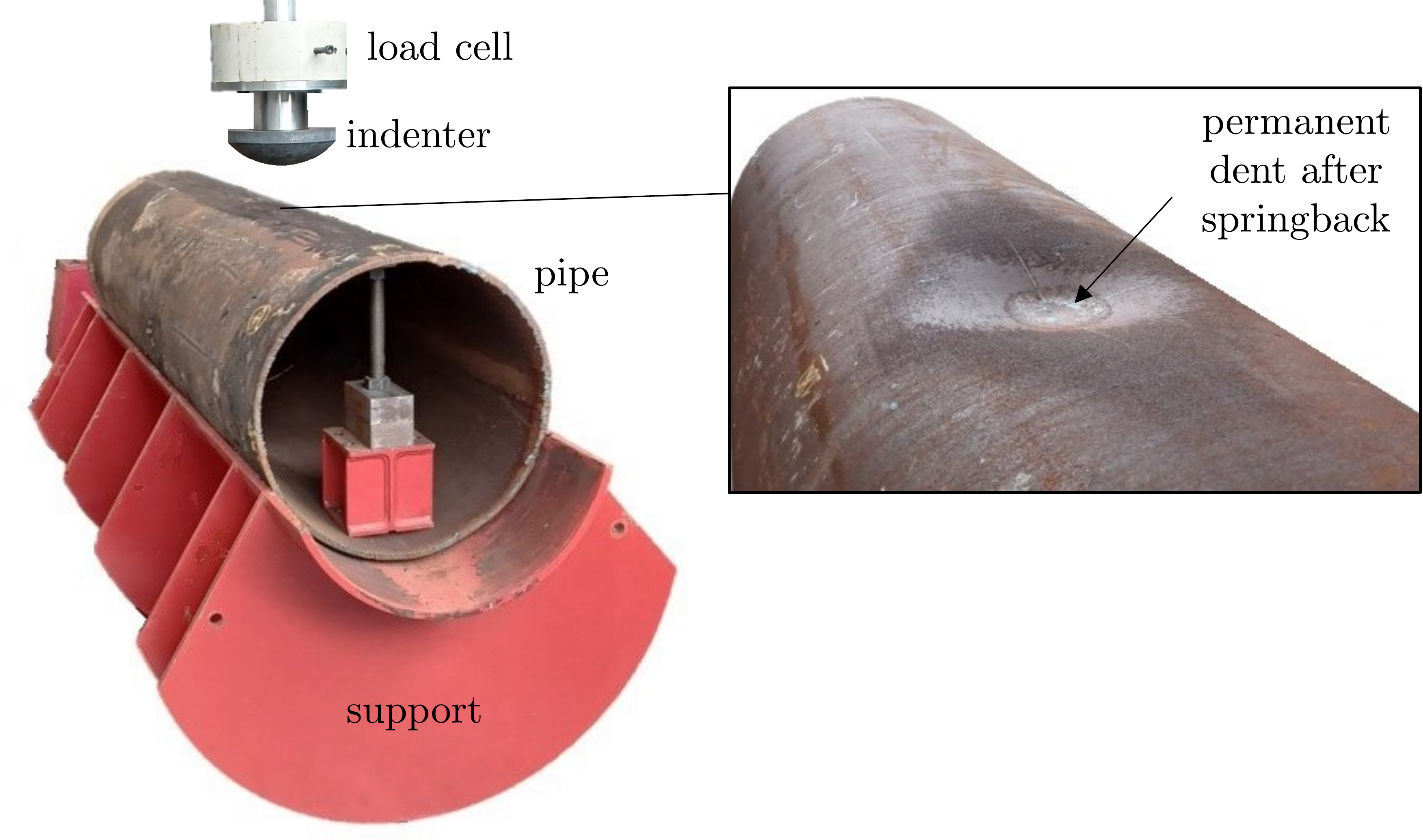} 
\caption{Full-scale pipeline dent testing: experimental setup, including a magnified image of the dent profile after removing the load.}
\label{F_Experiment_Setup}
\end{figure}

A full-scale pipeline of length $l=2$ m, internal diameter $d_0=733$ mm and thickness $h=14.75$ mm, with zero internal pressure, is quasistatically indented up to a depth of 102 mm and then allowed to spring back, leaving a permanent dent. The resulting dent profile is measured from the outer surface, and the load-displacement response is recorded with a suitable load cell, as shown in Fig. \ref{F_Experiment_Setup}, where the test setup is presented, as well as a close view of the dent profile. The pipe is supported at the bottom by a stiff circular-shaped guide that restricts vertical movement during indentation. As shown in Fig. \ref{F_Experiment_Setup}, vertical deformation is constrained on the edges of the pipeline. Denting is achieved by means of a circular indenter of 200 mm diameter. The required denting depth (102 mm) is achieved within 25 min., after which the indenter is removed to allow for springback. This experiment is then simulated by a 3D quasistatic, elastic-plastic simulation including contact, as described next.

\subsubsection{Numerical implementation}

\begin{figure}[tb]
\centering
\includegraphics[width=6 in]{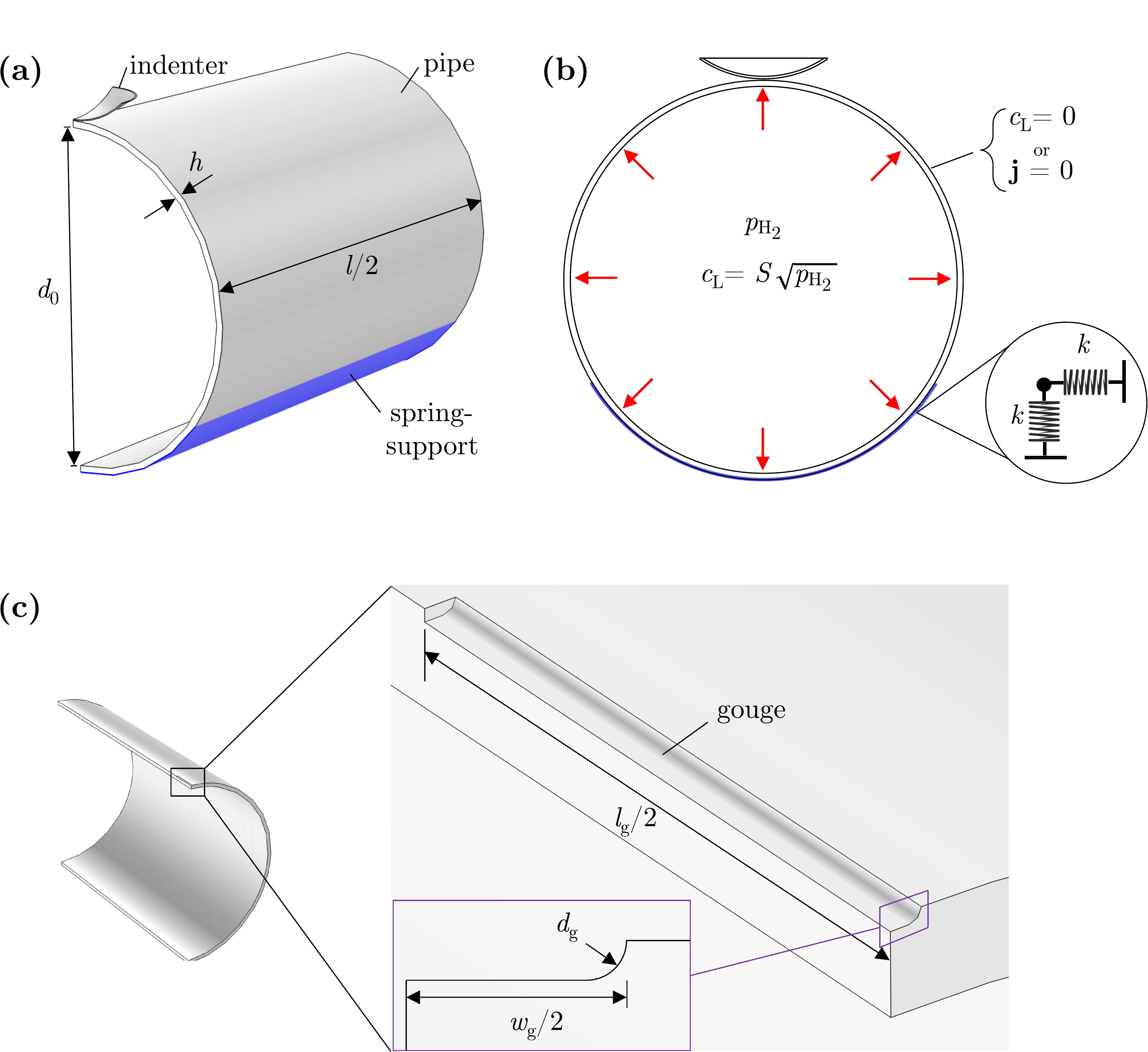}
\caption{Full-scale pipeline dent testing: details of the numerical model, including (a) the geometry and configuration of the underformed pipe and indenter, with the blue shaded area indicating the spring support; (b) 2D section illustrating the boundary conditions, including those relevant to the hydrogen transport problem, as assessed in Section \ref{Sec:Results}; and (c) Geometry and location of the gouge for the dent and gouge analysis. Only a quarter of the geometry is modelled, taking advantage of symmetry.}
\label{F_geom_BC}
\end{figure}

The numerical model mimics the dimensions and conditions of the full-scale experiment, as shown in Fig. \ref{F_geom_BC}. Taking advantage of symmetry, only a quarter of the pipe and indenter are simulated. The same model is used in Section \ref{Sec:Results} to deliver predictions in H$_2$ environments, with the presence of hydrogen being accounted for through the application of an inner pressure and by using Sievert's law to determine the associated lattice hydrogen concentration in the inner boundary. The elastic and plastic properties of the material are obtained from a uniaxial tension test conducted on a sample taken from the pipe along its transverse direction. This resulted in a Young's modulus of $E=187$ GPa, a Poisson's ratio of $\nu=0.3$, an initial yield stress of $\sigma_\text{y0}=473$ MPa and a strain hardening exponent of $N=0.093$.\\

Contact between the indenter and the pipe is modelled using the Augmented Lagrangian method. The contact constraint is formulated using the following Karush-Kuhn-Tucker (KKT) conditions
\begin{equation}
    g_{\mathrm{n}} \geq 0, \quad p_{\mathrm{n}} \leq 0, \quad 
    g_{\mathrm{n}} p_{\mathrm{n}} = 0, \text{ and} \quad 
    \dot{g}_{\mathrm{n}} p_{\mathrm{n}} = 0,
\end{equation}
where $g_{\mathrm{n}}$ is the normal gap function and $p_{\mathrm{n}}$ is the normal contact pressure. These conditions ensure that the contact forces are activated only when the surfaces are in contact ($g_{\mathrm{n}} \leq 0$) and vanish upon separation. The contact pressure is introduced via a penalty formulation, in which
\begin{equation}
    \lambda_{\mathrm{n}} = -k_\text{p} g_{\mathrm{n}} \quad \text{for} \quad g_{\mathrm{n}} < 0,
\end{equation}
with $\lambda_{\mathrm{n}}$ being the appropriate Lagrange multiplier, and $k_\text{p}$ being the penalty stiffness. Contact between the pipeline and the support is modelled by means of a spring base with a spring constant $k = 7.5 \times 10^{5} $ N/m. The spring support region is shown in Fig. \ref{F_geom_BC}(a). For clarity, a 2D view of the transverse plane at the pipeline's longitudinal midsection is also depicted in Fig. \ref{F_geom_BC}(b), where the inset shows the configuration of the spring support on the 2D plane (no spring resistance is considered along the longitudinal direction).\\

To assess the effect of the defect geometry on the deformation of a pipe, simulations are also conducted for the case of a dented pipe that also contains a gouge on the dented surface. This is shown in Fig. \ref{F_geom_BC}(c), with the gouge having a depth of $d_\text{g}=2$ mm, a length of $l_\text{g}=200$ mm and a width of $w_\text{g}=10$ mm. The 3D models of the pipe and indenter are discretised using 20-node second-order serendipity brick elements. Following a mesh sensitivity study, approximately 1 million degrees of freedom (DOFs) are used for the dented pipeline mesh and slightly over 2 million DOFs for the pipe containing a dent and a gouge. In both cases, the mesh is refined near the contact region.

\subsubsection{Full-scale verification results}

\begin{figure}[]
\centering
\includegraphics[width=6.1 in]{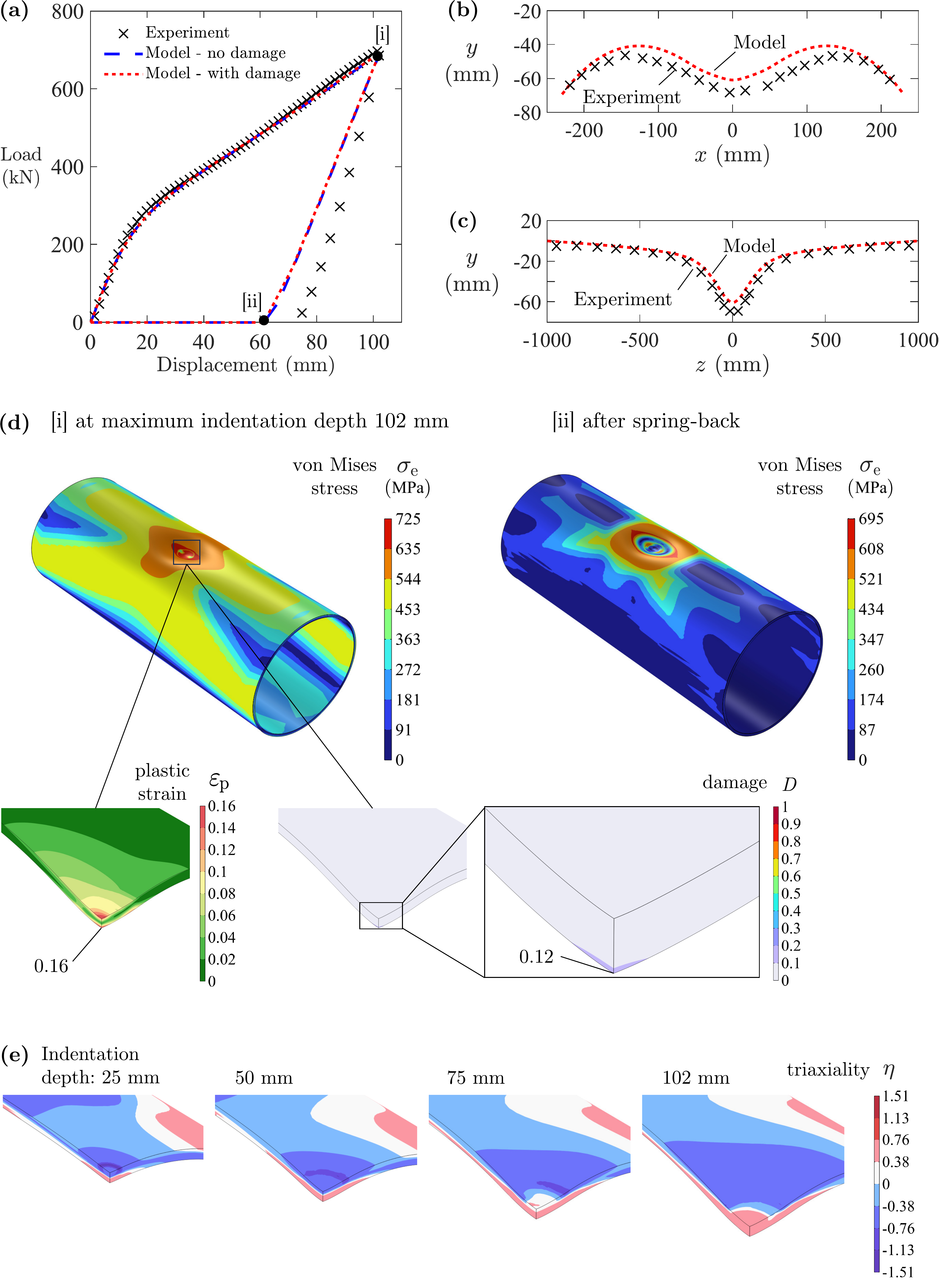} 
\caption{Indentation of a hydrogen-free pipe at zero internal pressure: (a) Comparison of the load versus displacement response obtained experimentally and with the numerical model; Comparison of experimentally measured and numerically predicted dent shapes at (b) the transverse midsection and (c) the longitudinal midsection; (d) distribution of von Mises stress, plastic strain and damage at [i] maximum indentation and [ii] after springback; and (e) distribution of triaxiality ($\eta=\sigma_\text{h}/\sigma_\text{e}$) around the dent at indentation depths from 25 mm up to 102 mm.}
\label{F_FIG_Result_load_disp}
\end{figure}

The load-displacement curve obtained from the simulation is compared with the experimental measurement in Fig. \ref{F_FIG_Result_load_disp}(a). The load exerted by the indenter is determined by integrating the vertical component of the contact pressure on the outer surface of the pipe. A very good agreement is attained, particularly during the loading phase. Simulations are run with and without the damage model. As shown in Fig. \ref{F_FIG_Result_load_disp}(a), damage has a negligible influence on the load versus displacement response. This is to be expected as denting by itself does not usually result in cracking, and no damage was observed in the experiments. Slight differences between the model and the experiment are attained during unloading: the simulation predicts a permanent dent depth of approximately 62 mm, in contrast to the 68 mm depth measured in the test. We speculate that this difference arises due to the idealisation of the pipe-support in the model as an ideal spring, compared to the complex polyvinyl chloride (PVC) support used in the experiment. In addition, the measured and simulated dent shapes at the transverse and longitudinal midsections of the pipe are compared in Fig. \ref{F_FIG_Result_load_disp}(b) and (c), respectively. The shape and width of the dent show excellent agreement between the experiment and the simulation. The results confirm that the model can accurately capture the mechanics of indentation and the dent shape.\\ 

% \subsubsection{Further discussion}
To gain insight into the local conditions near dents, contours of relevant fields are shown in Figs. \ref{F_FIG_Result_load_disp}(d) and (e). Specifically, the distributions of von Mises effective stress $\sigma_\text{e}$, equivalent plastic strain $\varepsilon_\text{p}$, and damage $D$ at the maximum indentation depth of 102 mm are presented. Significant levels of plastic deformation are observed, with the equivalent plastic strain attaining a peak value of $\varepsilon_\text{p}=0.16$. Plastic deformation levels are particularly high at the dent centre, in both the outer (contact region) and inner surfaces - see Fig. \ref{F_FIG_Result_load_disp}(d). This is also where the highest stresses are attained, with the maximum von Mises effective stress being equal to $\sigma_\text{e}=725$ MPa and attained at the dent surface. Damage, however, localises in the inner surface, where the stresses are tensile, and the stress triaxiality is highest (see Fig. \ref{F_FIG_Result_load_disp}(e)). The highest value of $D$ is small (0.12) and attained at the inner wall, opposite to the indenter contact point. The very low damage attained suggests that such levels of denting do not result in significant microcracking, even locally. This is in agreement with experimental observations.\\ 

The distribution of residual von Mises stress after springback is illustrated in subfigure [ii] of Fig. \ref{F_FIG_Result_load_disp}(d). As expected, the dented pipe partially recovers its elastic strain during unloading, thereby releasing a significant portion of the stress away from the dent. However, the peak von Mises effective stress remains high at 695 MPa close to the dent. The distributions, peak values, and locations of plastic strain and damage remain unchanged after springback due to their irreversible nature.\\

\begin{figure}[ht!]
\centering
\includegraphics[width=6 in]{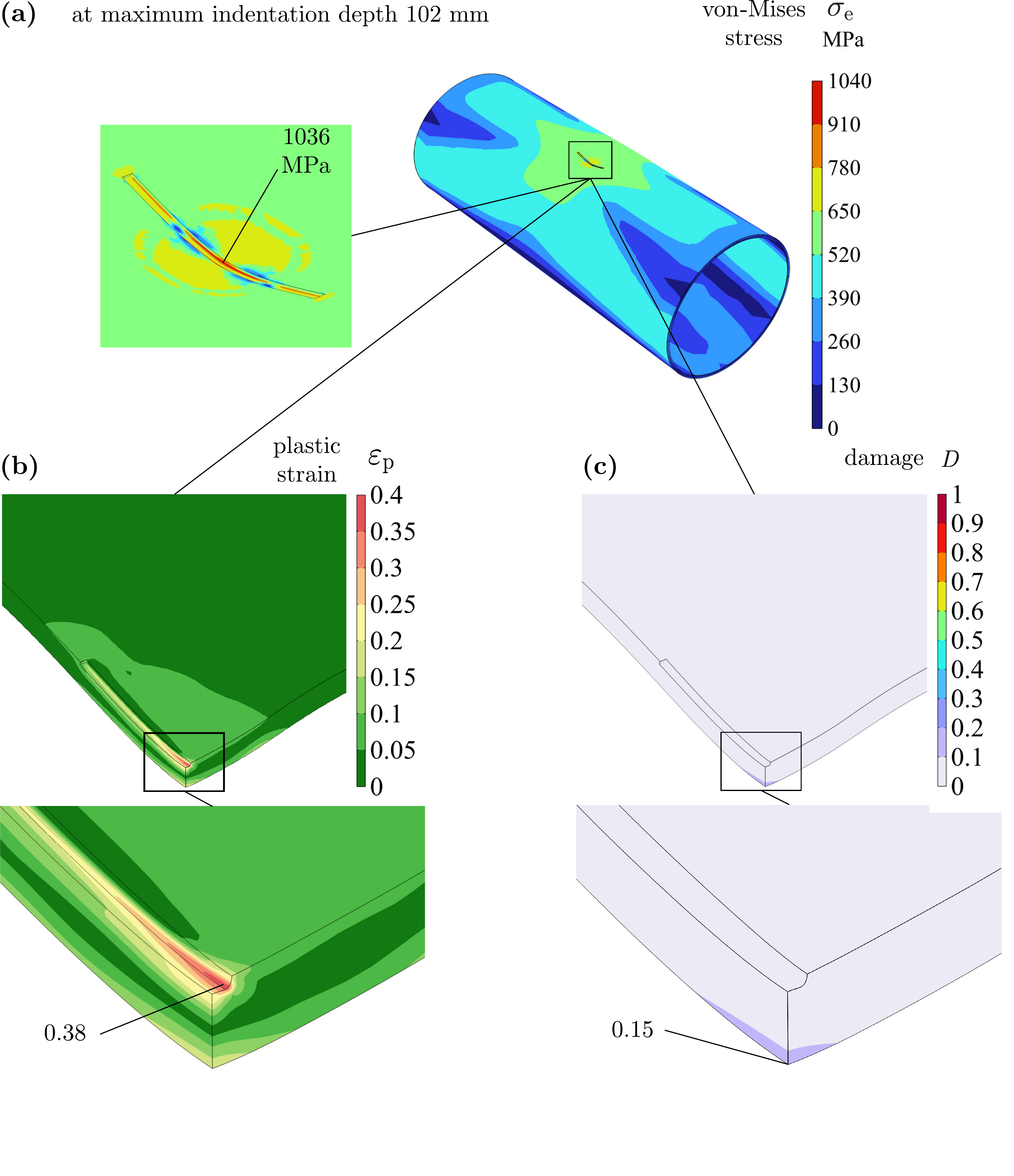}
\caption{Indentation of a hydrogen-free pipe at zero internal pressure \emph{with a gouge}; distribution of (a) von Mises effective stress $\sigma_\text{e}$, (b) equivalent plastic strain $\varepsilon_\text{p}$, and (c) damage $D$, in the vicinity of the dent region, at a maximum indentation depth of 102 mm.}
\label{F_result_ep_D_pass_G}
\end{figure}

Finally, let us consider the case of a pipeline undergoing denting and containing a gouge. The results are shown in Fig. \ref{F_result_ep_D_pass_G}. The gouge acts as a stress concentrator, raising the maximum $\sigma_\text{e}$ value to 1036 MPa, a 43\% increase relative to the gouge-free pipe (see Fig. \ref{F_result_ep_D_pass_G}(a)). Correspondingly, the maximum equivalent plastic strain reaches 0.38, more than double that of the gouge-free dent. Despite these localised rises, the maximum damage remains limited to 0.15 at the inner wall of the dent's centre, while the gouge itself shows almost no damage, see Fig. \ref{F_result_ep_D_pass_G}(b) and (c). The low damage in the gouge is attributed to the low stress triaxiality in the gouged region during indentation, which prevents damage accumulation in compressive regions.

\section{Analysis of a hydrogen transport pipe with external defects} 
\label{Sec:Results}

\begin{figure}[bt]
\centering
\includegraphics[width=6 in]{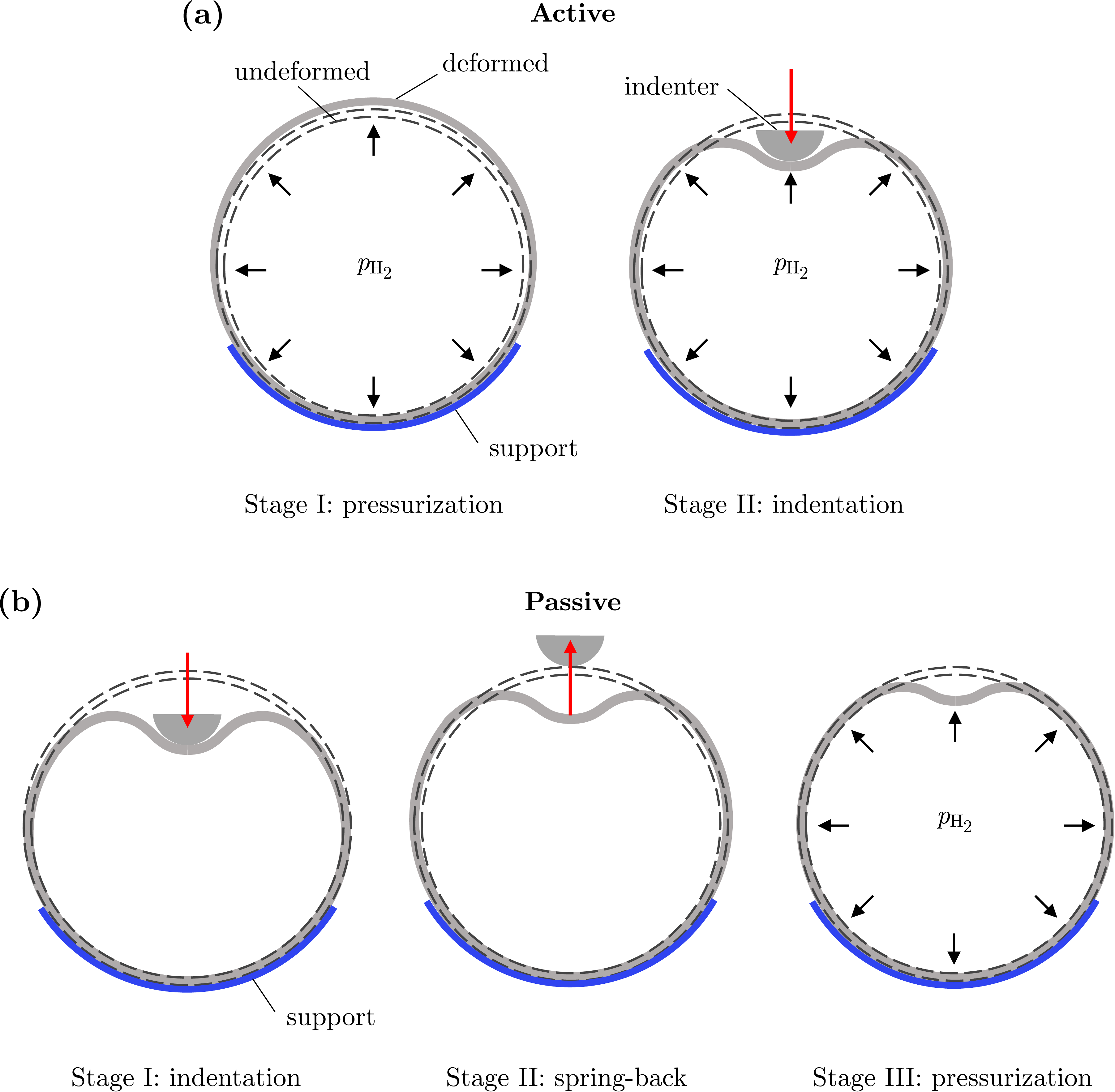}
\caption{Scenarios and modelling stages considered during (a) active indentation of a H$_2$ pressurised pipe, and (b) passive indentation, whereby a dented hydrogen-free pipe is exposed to H$_2$ pressure.}
\label{F_Active_Passive}
\end{figure}

Having validated the hydrogen and mechanical predictive capabilities of the model, we proceed to use it to shed light on the structural integrity of hydrogen transport pipelines containing dents and gouges. The coupled deformation-diffusion-damage simulations consider two main scenarios according to the sequence of hydrogen ingress and indentation: (see Fig. \ref{F_Active_Passive})

\begin{enumerate}[label=(\roman*)]

\item \textit{Active}: a pressurised, hydrogen-charged pipe undergoes indentation. Hence, there is an existing distribution of lattice and trapped hydrogen along the pipe's wall thickness, and the large strains resulting from the denting process result in trap creation (higher dislocation density) and hydrogen re-distribution. The sequence of modelling steps to simulate damage in an active dent, illustrated schematically in Fig. \ref{F_Active_Passive}(a), is as follows:  

\begin{itemize}
   \item Stage I: The pipe is first subjected to an internal hydrogen pressure $p_{\text{H}_2}$ until a steady-state hydrogen concentration is established across the wall thickness.  

   \item Stage II: A rigid spherical indenter is then pressed into the pipe’s external surface up to a prescribed indentation depth, inducing local plastic straining, which brings additional trap sites and damage.  
\end{itemize}

\item \textit{Passive}: a hydrogen-free pipe is indented, then allowed to spring back, and later exposed to hydrogen through internal pressurisation. In this case, the dent serves as a geometric imperfection that redistributes stresses and hydrogen concentration during re-rounding due to the application of an inner pressure, but hydrogen is not present during the dent creation process. The simulation sequence for the passive case, illustrated in Fig. \ref{F_Active_Passive}(b), is as follows:  

\begin{itemize}
   \item Stage I: A hydrogen-free pipe is indented up to a chosen depth using a rigid indenter.
   
   \item Stage II: The indenter is removed to allow elastic springback to an equilibrium dent depth. No hydrogen is present at this point. 

   \item Stage III: The dented pipe is subsequently subjected to a quasi-static rise in internal hydrogen pressure until failure. Hydrogen diffusion, damage in the wall, and trap density evolve concurrently with the stress redistribution and re-rounding of the dent.  
\end{itemize}

\end{enumerate}

In both active and passive scenarios, the modelling framework involves two strongly coupled components: (i) the mechanical indentation process, which dictates the local plastic strain, dislocation density distribution, and damage; and (ii) hydrogen diffusion and trapping, which significantly depend on the evolving stress and strain fields. We capture the exposure to a hydrogen pressure $p_{\text{H}_2}$ by using Sievert's law, which relates the lattice hydrogen concentration at the inner surface of the pipeline with $p_{\text{H}_2}$ and the material solubility $S$,
\begin{equation}
c_{\text{L}}^0 = S\sqrt{p_{\text{H}_2}} \,\, ,
\label{Eq:Sievert}
\end{equation}

\noindent with $S$ = 0.55 mol/(m$^3\sqrt{\text{MPa}})$ being the solubility of ferritic steel \cite{Martin2020}. Hydrogen is free to leave the pipeline at the outer surface, as captured through a Dirichlet boundary condition of zero lattice hydrogen concentration at the outer surface or $c_\text{L}^\text{out} = 0$. This boundary condition is likely to be the closest to reality, given the size of the hydrogen atom, and thus the one adopted majoritarily in this work. However, it does not reflect the most conservative scenario. In gas pipelines, hydrogen egress can be restricted by coatings, corrosion products or cathodic protection. Hence, in line with our aim to establish the most conservative scenario, a set of calculations is also conducted with the natural, no-flux boundary condition $\textbf{j} = -D_\text{L}\nabla c_\text{L}=0$ ($\mathbf{j}^\text{out} =0$), which represents the other limiting case, where the outer surface is impermeable to hydrogen. These boundary conditions are shown schematically in Fig. \ref{F_geom_BC}(b). The hydrogen transport equation (\ref{Eq:modFick}), considering Eq. (\ref{Eq:HTransportmult}) for the trap evolution, and the balance of linear momentum, Eq. (\ref{Eq:MomentumBalance}), considering the degradation law Eq. (\ref{Eq:DamageEvolution}), are solved in a fully coupled manner. 

\subsection{Active damage}

Let us begin our analysis by considering the case of active indentation. First, a hydrogen pressure of ${p_{\text{H}_2}} = 10$ MPa ($\sim$100 bar) is applied. The mechanical deformation at this internal pressure remains within elastic limits. The hydrogen boundary condition at the inner wall is considered by prescribing the corresponding lattice hydrogen concentration, as per Eq. (\ref{Eq:Sievert}), and the outer surface boundary condition is chosen accordingly ($c_\text{L}^\text{out} = 0$ or $\mathbf{j}^\text{out} = 0$), as explained above. The hydrogen transport simulation is conducted until a steady state distribution of hydrogen is achieved. Then, quasistatic indentation is applied up to a depth of 102 mm, a magnitude of practical relevance. By selecting a quasistatic indentation rate of $10^{-7}$ m/s, we ensure that lattice hydrogen and stress-driven fluxes remain at steady state throughout the indentation process, providing a conservative estimate of hydrogen distribution and damage. The results obtained are assessed by considering first the redistribution of hydrogen during denting, and evaluating damage evolution subsequently. 

\subsubsection{Hydrogen distribution around dent}

\begin{figure}[H]
\centering
\includegraphics[width=6.4 in]{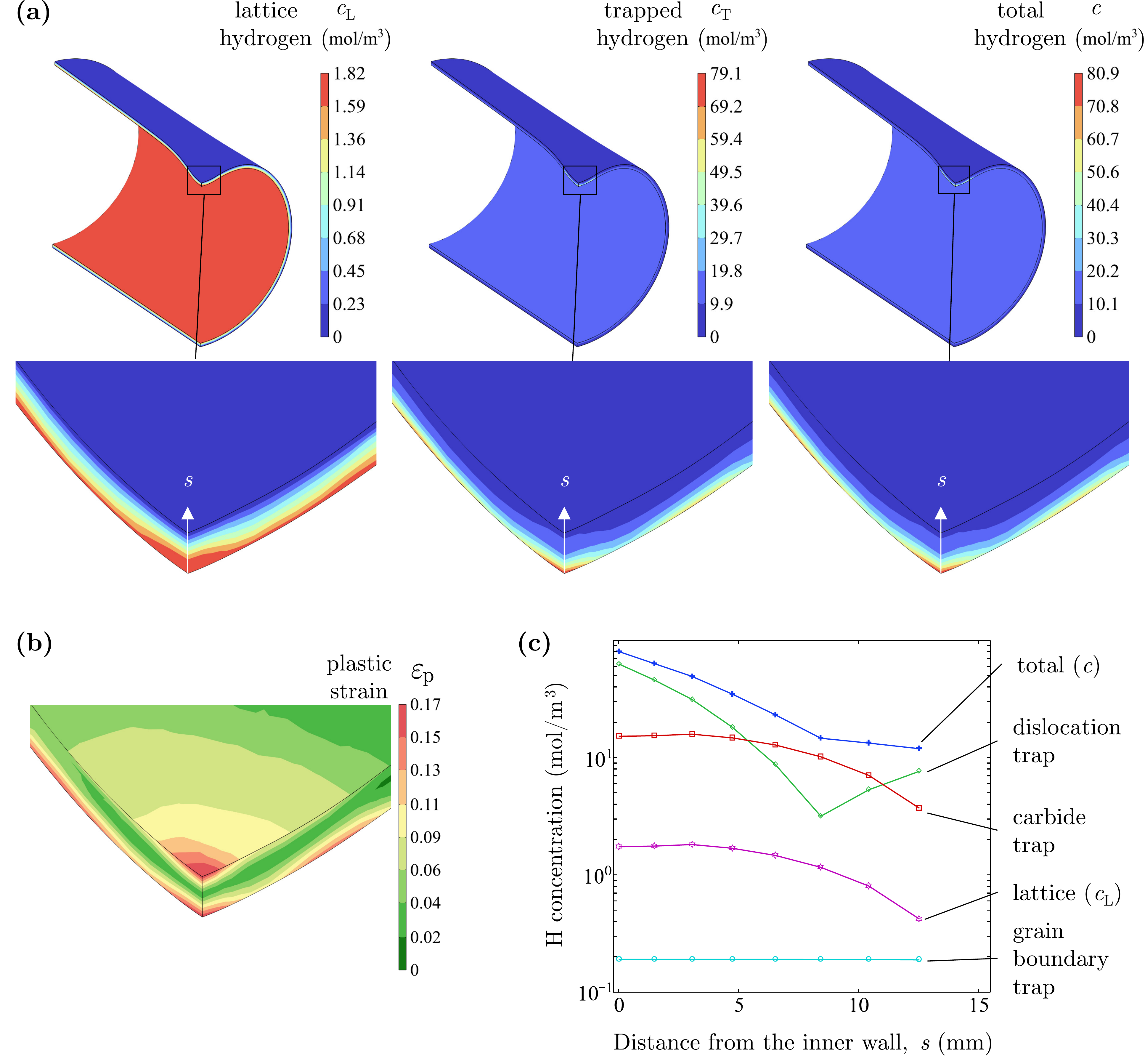}
\caption{Hydrogen distribution in a gouge-free pipe carrying H$_2$ at a pressure of 10 MPa and undergoing active indentation (102 mm depth): (a) contours of lattice, trapped, and total hydrogen concentration, with a local coordinate system ‘$s$’ being defined across the pipe’s wall at the centre of the dent; (b) contours of plastic strain around the dent’s centre; and (c) distribution of hydrogen in lattice and various trap sites across the pipe’s wall at the dent's centre, using the previously defined local coordinate system.}
\label{F_FIG_Act}
\end{figure}

The hydrogen distributions obtained in the active case at the maximum indentation depth (102 mm) are given in Fig. \ref{F_FIG_Act}. Results are shown for a pipe without a gouge, but these are qualitatively similar to those obtained in the case of a gouged pipe. Contours of lattice ($c_{\text{L}}$), trapped ($\sum_\text{i} c_{\text{T}}^{\text{i}}$), and total ($c $ = $ c_{\text{L}}$+$\sum_\text{i} c_{\text{T}}^{\text{i}}$) hydrogen concentrations are given in Fig. \ref{F_FIG_Act}(a). The $c_\text{L}$ distribution reveals the expected trend, changing progressively from the inner hydrogen content to its zero value at the outer surface, where hydrogen atoms are allowed to exit. A notable feature in the $c_\text{L}$ distribution is the local accumulation near the inner wall at the dent centre, as highlighted in the magnified inset. This accumulation arises from the hydrostatic stress–driven flux, which enhances lattice hydrogen migration into this region. As described by Oriani's equilibrium, Eq. (\ref{Eq:Oriani}), this elevated lattice occupancy near the dent directly promotes trap occupancy; note the contours of trapped hydrogen closely following the contours of lattice hydrogen. Trapped hydrogen concentration attains a maximum of 80 mol/m$^3$ at the inner wall of the dent centre — almost 40 times higher than the peak lattice hydrogen concentration. A higher trapping content is a common feature in bcc steels due to their low lattice solubility, and this is further exacerbated here by the creation of numerous dislocation trap sites due to the large plastic straining taking place, see Fig. \ref{F_FIG_Act}(b).\\

The relative contributions of individual trap types at the dent centre are examined in Fig. \ref{F_FIG_Act}(c), where hydrogen concentrations across the wall thickness are plotted against the local coordinate $s$ (defined in Fig. \ref{F_FIG_Act}(a)). The hydrogen trapped in carbides follows an almost linear relationship with lattice concentration, as expected from Eq. (\ref{Eq:Oriani}). Grain boundary traps, characterised by their deep binding energy of $-50$ kJ/mol, saturate (i.e. $\theta_\text{T}^\text{gb} \approx 1$) at $c_\text{L} = 0.05$ mol/m$^3$ (see Fig. \ref{F_Energy_Landscape}(b)), yielding a nearly constant concentration profile across the thickness, before the outer surface is approached. Due to their lower trap density, the hydrogen content in grain boundaries is smaller relative to other trap types. Dislocations dominate the trapped hydrogen population, reaching nearly 73 mol/m$^3$ at $s=0$. This peak reflects the elevated dislocation density induced by severe plastic straining at the dent centre. A local minimum in the hydrogen content trapped at dislocations appears at the mid-thickness ($s=7.4$ mm), consistent with the plastic strain distribution, as shown in Fig. \ref{F_FIG_Act}(b). All trap contents reduce to 0 at the outer surface, since $c_\text{L}=0$ there, and hydrogen in trap sites is assumed to be in equilibrium with that in lattice sites, a sensible assumption \cite{garcia2024tds}.  

\subsubsection{Damage evolution}

\begin{figure}[H]
\centering
\includegraphics[width=6.2 in]{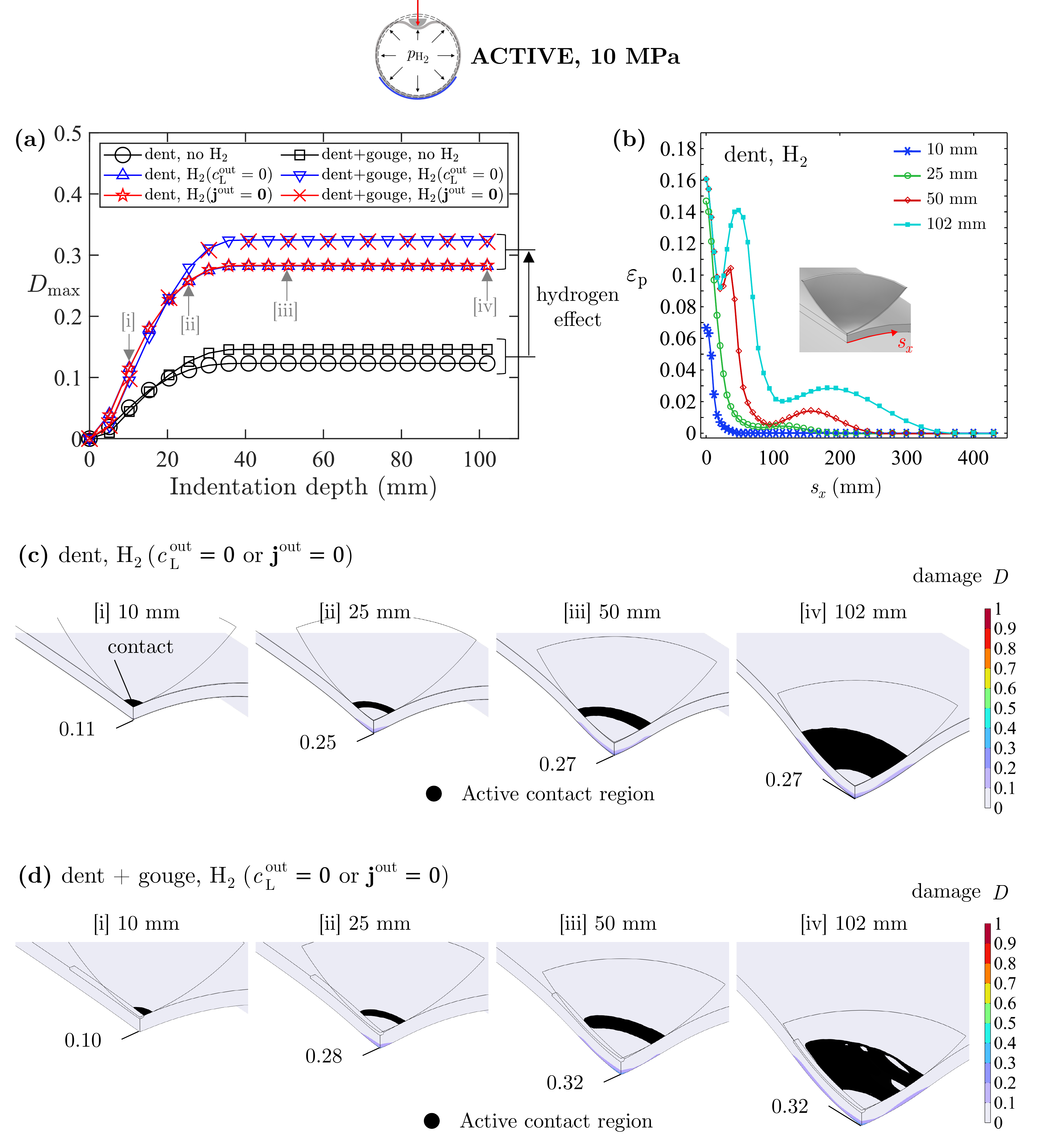}
\caption{Damage evolution during active indentation in gouged and gouge-free pipes containing H$_2$ at a 10 MPa pressure. (a) Evolution of the maximum damage $D_{\text{max}}$ in the pipe, which is attained at the inner wall of the dent’s centre, as a function of indentation depth. (b) Evolution of the equivalent plastic strain along the pipe inner surface, at various indentation depths in a gouge-free pipe. Evolution of the contact regions between indenter and pipe, and damage at various indentation depths in a (c) gouge-free and in (d) a gouged pipe.}
\label{F_Act_Damage}
\end{figure}

To quantify the impact of hydrogen-induced degradation on pipeline integrity, we choose to plot $D_\text{max}$, the maximum value of $D$ attained in the entire model, as a function of the indentation depth. The results are given in Fig. \ref{F_Act_Damage}(a), spanning the cases of a dented pipe without a gouge and a pipe containing both a dent and a gouge, with hydrogen ($p_{\text{H}_2}=10$ MPa) and without it ($p_{\text{H}_2}=0$). In addition, the two scenarios in terms of hydrogen behaviour in the outer surface are considered: free egress ($c_\text{L}^\text{out}=0$) and impermeable outer surface ($\textbf{j}^\mathrm{out}=0$). In both hydrogen and hydrogen-free scenarios, the maximum level of damage is attained at the inner wall of the dent centre, due to the high levels of plastic strain, triaxiality, and hydrogen content reached there. Consistent with the results shown in Section \ref{Sec:Ind_H_free}, the maximum level of damage is very limited in the hydrogen-free scenarios, with the maximum being very localised and between $D = 0.12$ (for the dent without a gouge) and $D = 0.15$ (for the dent with a gouge). Hydrogen elevates those numbers roughly two-fold, to $D = 0.27$ and $D = 0.32$, respectively. Nevertheless, this degree of damage, which remains mainly localised in a very small region (see Figs. \ref{F_Act_Damage}(c) and (d)), is still very low and has a negligible effect on the load-carrying capacity of the pipe. In other words, active denting under conditions that do not result in cracking in natural gas pipelines is unlikely to result in cracking in pipelines carrying H$_2$ at pressures on the order of 10 MPa. This conclusion is attained irrespectively of the outer surface scenario, with predictions showing almost no sensitivity to the outer boundary condition ($c_\text{L}^\text{out} = 0$ vs $\mathbf{j}^\text{out} = 0$).\\

The $D_{\text{max}}$ versus indentation depth results also reveal that, in all cases, the maximum damage attained saturates once the indentation depth exceeds 40 mm. This trend is attributed to the saturation of plastic strain at the dent centre beyond this depth, while plastic strain continues to accumulate in surrounding regions. This is quantified in Fig. \ref{F_Act_Damage}(b), where the equivalent plastic strain distribution along the arc $s_x$ is shown for the gouge-free pipe case, with the results being representative of the dent with a gouge scenario too. The curves show that the maximum level of $\varepsilon_\text{p}$ is attained at $s_x=0$ and saturates at a value of 0.16 for indentation depths beyond 40 mm, with plastic straining continuing to increase away from the dent. The evolution of plastic strain is governed by the changing contact conditions between the indenter and the pipe. Simulations reveal that the indenter partially loses contact with the outer wall once the indentation depth reaches 25 mm, as illustrated by the contact regions (in black) in Fig. \ref{F_Act_Damage}(c) for the gouge-free pipe and in Fig. \ref{F_Act_Damage}(d) for the gouged pipe. This loss of contact reduces deformation at the dent centre and slows further plastic strain accumulation there, consistent with the plateau observed between 25 mm and 50 mm in Fig. \ref{F_Act_Damage}(b). New contact regions are formed at a distance from the dent centre, as shown in Fig. \ref{F_Act_Damage}(c)–(d), resulting in additional regions of plastic strain localisation and damage. This is a similar scenario to that of the small punch test \cite{martinez2016damage}, suggesting that there would be an indentation depth (far beyond the depths expected in natural gas pipelines) at which cracking would occur, with cracks originating in a circular band slightly away from the dent (or punch) centre. 

\subsection{Passive damage}

Since dents are known to be already present in the natural gas pipeline network, it is necessary to understand how they interact with hydrogen and potentially limit H$_2$ transmission service conditions. To assess this, we simulate the behaviour of a dented pipeline under rising pressure and determine at what point it fails. This is conducted under both the case of an inner natural gas pressure and the case of H$_2$ pressure, for comparison. Also, as in the active case, two scenarios are considered in terms of hydrogen egress from the outer pipe surface: (i) free hydrogen desorption, captured through a $c_\text{L}^\text{out}=0$ boundary condition, and (ii) completely hindered hydrogen desorption, captured through a $\mathbf{j}^\mathrm{out}=0$ boundary condition. To be on the conservative side, the internal pipeline pressure is raised quasi-statically, at a low rate (0.001 MPa/s), allowing hydrogen to re-distribute, until either (i) $D=1$ at any integration point in the model, or (ii) the pipe loses stiffness due to extensive plastic flow. The critical pressure attained when one of these two conditions is met is taken as the pipe's burst pressure.\\  

\begin{figure}[ht!]
\centering
\includegraphics[width=6.5 in]{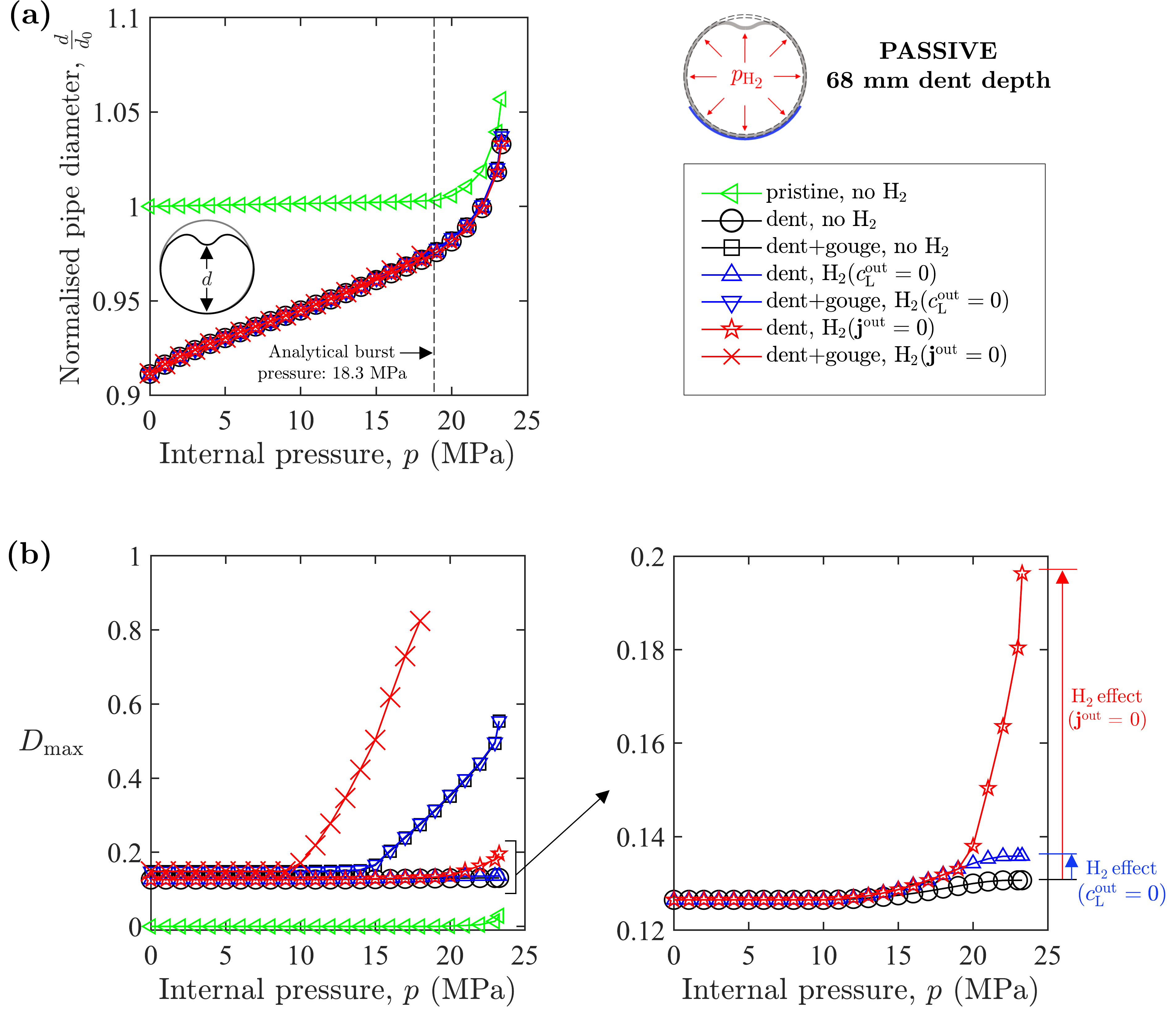}
\caption{Analysis of the integrity of pipelines in the passive condition, where denting precedes H$_2$ exposure: (a) Evolution of the normalised deformed inner diameter $d/d_0$ as a function of rising internal pressure, with the analytically-estimated burst pressure indicated with a dash line; and (b) evolution of the maximum damage attained in the pipeline $D_{\text{max}}$ as a function of internal pressure, with the inset providing a detailed view of the effect of H$_2$.}
\label{F_Pass}
\end{figure}

The results obtained are shown in Fig. \ref{F_Pass}. In total, seven scenarios are considered: (i) a pristine pipeline without hydrogen and external defects, (ii) a hydrogen-free pipeline with a dent, (iii) a hydrogen-free pipeline with a dent and a gouge, (iv) a pipeline with a dent becoming exposed to H$_2$, (v) a pipeline with a dent and a gouge becoming exposed to H$_2$,(vi) a dented pipeline exposed to H$_2$, but where hydrogen can not leave the outer surface, and (vii) a pipeline with a dent and a gouge becoming exposed to H$_2$, but where hydrogen can not leave the outer surface. The dent is assumed to be 68 mm deep, the geometry adopted corresponds to that obtained after the spring-back of the 102 mm dent in the validation case in Section 3.3.3, and the gouge dimensions are given in Fig. \ref{F_geom_BC}(c). For the hydrogen-containing environment, a lattice hydrogen concentration is prescribed at the inner surface of the wall, corresponding to the gas pressure considered, as per Sievert's law (\ref{Eq:Sievert}). Consider first Fig. \ref{F_Pass}(a), where the normalised pipe diameter $d/d_0$ is plotted against the incrementally increased internal pressure for a defect- and hydrogen-free pristine pipeline. The curve shows two regimes: a negligible diameter increase up to 20 MPa, followed by a rapid expansion beyond this pressure, when the entire pipeline cross-section undergoes plastic deformation, eventually leading to a numerical instability, as material hardening becomes insufficient to counteract the geometric softening arising from wall thinning and the resulting stress redistribution. The burst pressure is taken to be the one at the onset of the rapid diameter expansion, $\sim$20 MPa. This predicted value agrees well with existing analytical estimates of burst pressure; e.g., Barlow's formula \cite{BARLOW1836} gives a burst pressure of $P_\text{max} = 2h\sigma_{\text{y0}}/(d_0 + 2h) = 18.3 \text{ MPa}$. Consider now the result obtained for the hydrogen-free but defect-containing pipeline. The cases with and without a gouge beneath the dent overlap in Fig. \ref{F_Pass}(a), signifying nearly no effect of the external gouge on the pipe's burst pressure or failure mechanism. Different to the pristine pipe, the normalised diameter is below 1 at zero pressure, a direct consequence of the residual dent depth reducing the undeformed inner diameter $d_0$. Yet again, two clear regimes are observed: a relatively smaller increase in the diameter with internal pressure up to a pressure of 20 MPa, as the dent reshapes, followed by a rapid increase until a numerical instability arises due to plastic collapse. I.e., the failure mechanism and failure pressure observed for the defect-free pipeline also apply to this case.\\

The exposure to H$_2$ brings negligible changes in the pipeline diameter versus internal pressure behaviour if hydrogen is allowed to leave freely through the outer surface. As shown in Fig. \ref{F_Pass}(a), results reveal the same burst pressure and failure mechanism (plastic collapse), with the curves overlapping with those obtained for the hydrogen-free cases. However, for the case (vii) where a pipeline with a dent and a gouge becomes exposed to H$_2$ with the outer surface being impermeable to hydrogen, the pipe fails due to the crack initiation ($D=1$) at the gouge, at $p_{\text{H}_2}\sim 18$ MPa.\\

\begin{figure}[]
\centering
\includegraphics[width=6.2 in]{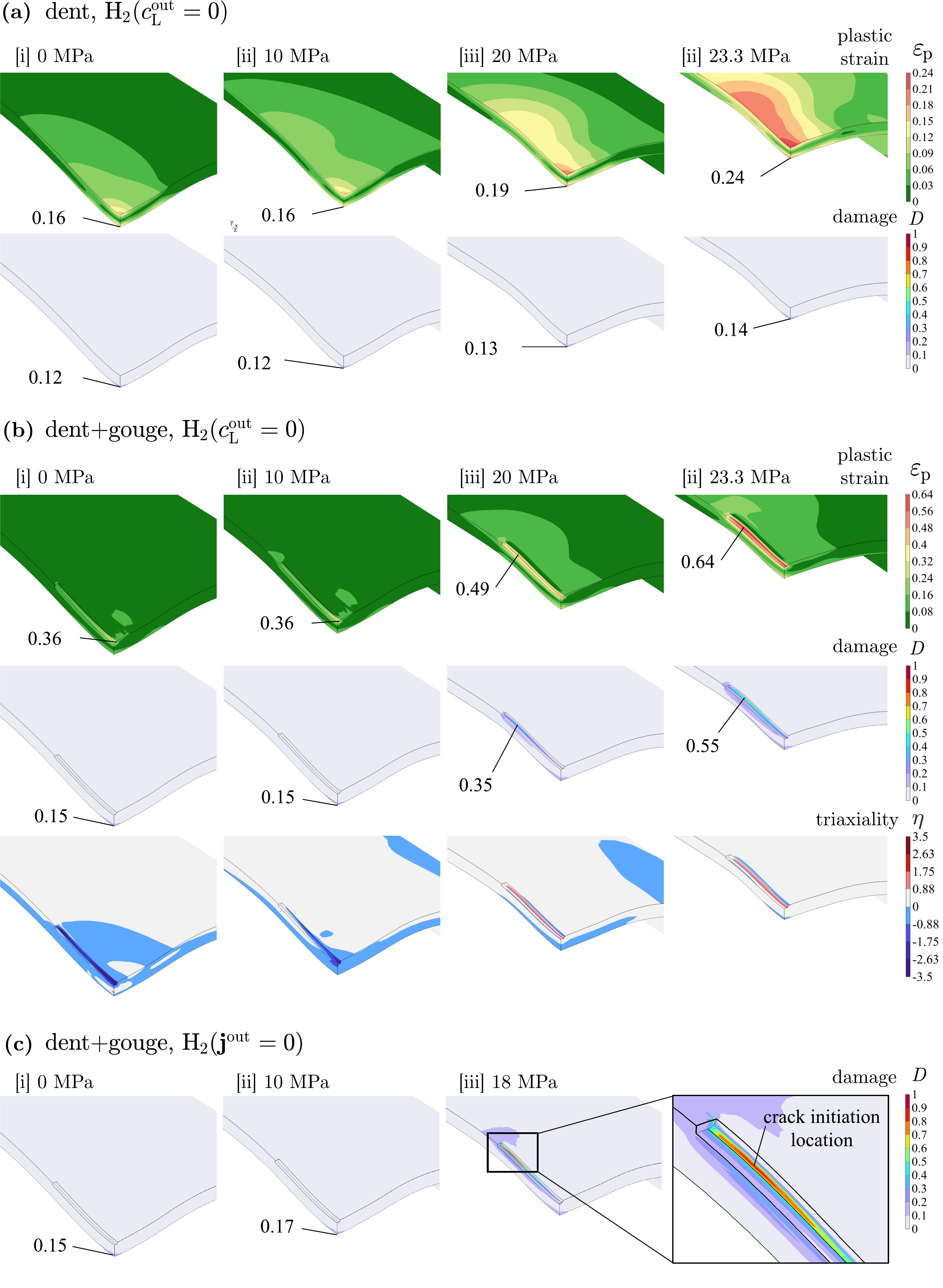}
\caption{Contours of equivalent plastic strain $\varepsilon_\text{p}$ and damage $D$ as a function of the internal hydrogen pressure for a pipeline that has undergone denting prior to H$_2$ exposure (i.e., \emph{passive} denting), and for the cases of: (a) a pipeline containing a dent but no gouge, (b) a pipeline containing a dent and a gouge, and (c) a pipeline containing a dent and a gouge with an outer surface impermeable to hydrogen.}
\label{F_Pass_Damage}
\end{figure}

The highest value of $D$ attained in the pipeline is plotted against the internal pressure in Fig. \ref{F_Pass}(b), to provide insight into the development of local damage. The defect- and hydrogen-free pristine curve shows negligible local damage, with $D_\text{max}$ remaining below 0.05 until the plastic collapse instability is triggered. However, the presence of external defects does raise local damage, and geometry is seen to play a significant role. As shown in Fig. \ref{F_Pass}(b), the dented pipe begins with a level of damage just below 0.15, as a result of the indentation process. In the absence of a gouge, this level of damage remains nearly constant up to the plastic collapse event, with hydrogen having a very small effect, as shown in the contour plots of Fig. \ref{F_Pass_Damage}(a). However, the presence of a gouge brings a notable increase in local damage. For the cases of no hydrogen, and hydrogen being allowed to leave the outer surface of the pipeline, a nearly identical behaviour is observed, characterised by three stages; local damage remains nearly constant up to a pressure of 15 MPa, followed by a sharp and nearly linear increase with applied pressure until the sudden rise in $D_{\text{max}}$ associated with the plastic collapse regime. This suggests a competition between a local damage mechanism and a global plastic collapse event, albeit the latter is eventually responsible for pipeline failure under these conditions. This sudden rise in local damage at a pressure below the plastic collapse-driven burst pressure corresponds to a shift in the critical damage location---from the dent’s inner wall to the gouge base---where high triaxiality drives rapid damage growth, even though this region has a low hydrogen content. This is shown in Fig. \ref{F_Pass_Damage}(b), where contours of damage and equivalent plastic strain are presented as a function of the internal pressure. It can be seen that the dented pipe, even in the presence of H$_2$, remains at a low level of damage, with damage being constrained to a very small region. However, in the case of a pipeline containing a dent with a gouge, pressures above $\sim$15 MPa trigger a strain/stress redistribution that results in high levels of triaxiality near the gouge tip and a notable increase in localised damage relative to the gouge-free dented pipeline scenario. Nevertheless, this local damage is not enough to initiate macroscopic cracking (achieved when $D=1$), as plastic collapse is attained before. However, if hydrogen is not allowed to leave the pipe, then the damage in the gouge rises faster, driven by a higher concentration of hydrogen at the gouge's outer surface. The rising damage initiates a crack ($D=1$) at the base of the gouge when $p_{\text{H}_2} > 18$ MPa. As shown in Fig. \ref{F_Pass}(b), this crack nucleation event is preceded by a notable rise of the maximum degree of damage, which initiates at a pressure of around 10 MPa. Fig. \ref{F_Pass_Damage}(c) shows the damage contours near the cracking pressure, showing clear localisation near the gouge. As such, the most detrimental scenario is found to be a pipeline that has undergone denting prior to H$_2$ exposure, containing a dent with a gouge and where hydrogen egress from the outer surface is being impeded.

\section{Conclusions} 
\label{Sec:Conclusions}

Experiments and modelling have been combined to shed light on the role that external defects, such as dents and gouges, can play in compromising the structural integrity of hydrogen transmission pipelines. A novel coupled deformation-diffusion-damage model has been developed that integrates a (i) multi-trap hydrogen transport formulation with stress-assisted diffusion and plasticity-driven trap evolution, (ii) a finite strain plasticity constitutive model, and (iii) a hydrogen- and triaxiality-dependent damage mechanics model. Each element of the model is rigorously calibrated against permeation, mechanical, and dedicated full-scale pipeline indentation experiments on pipeline steels. The model is shown to accurately capture hydrogen transport, damage, and localised pipeline deformation. The validated model is then employed to resolve the interplay among hydrogen, external defects and pipeline integrity, with a focus on two configurations: \emph{active} denting, whereby a pipeline transporting H$_2$ is indented, and \emph{passive} denting, where the indentation has occurred before H$_2$ exposure. Two limit scenarios are considered in terms of the chemical boundary conditions at the outer surface: (i) $c_\text{L}^\text{out} = 0$ resembling bare steel, and (ii) $\textbf{j}^\mathrm{out}=0$ resembling a surface perfectly coated by an impenetrable hydrogen barrier. The main findings include:

\begin{itemize}

    \item In active indentation, although hydrogen brings a two-fold increase in localised damage, the levels of damage predicted remain very low, with damage remaining localised at the inner wall of the dent's centre and no macroscopic cracking being predicted. The presence of a gouge caused negligible damage rise, as the area beneath the gouge contained compressive stresses. Importantly, the degree of damage attained was mainly insensitive to: (i) indentation depth, beyond 40 mm and up to 102 mm, and (ii) hydrogen content at the outer surface of the pipe (i.e., same findings for both free egress and impermeable barrier scenarios).

    \item In the passive case, the response predicted is very sensitive to the outer pipeline condition. If hydrogen can exit the pipeline (arguably the most common scenario), failure occurs due to plastic collapse at the same burst pressure in all cases, irrespective of the presence of dents/gouges or H$_2$. Hydrogen and defect geometry can increase local damage levels, with geometry dominating over H$_2$ effects, but these remain low and below the cracking threshold. However, when the pipeline is assumed to be enclosed within a fully hydrogen impermeable barrier ($\textbf{j}^\mathrm{out}=0$ condition), then hydrogen-assisted cracking can be observed, at a pressure slightly smaller than the burst pressure, but only if the dent contains a gouge.

    \item The most detrimental scenario is found to be the one where pipelines that have been dented before exposure to a rising H$_2$ pressure (\emph{passive} denting) contain a dent with a gouge and have a surface state that completely blocks hydrogen egress. In passive denting, damage is most serious on the outer surface, at the base of the gouge, and hence very susceptible to the hydrogen exit condition, with the permeation-free case showing almost no impact of H$_2$ on damage. In \emph{active denting}, the dent/gouge base is under compression and damage localises in the inner surface, where hydrogen content is highest and nearly insensitive to the outer surface condition; however, H$_2$ is not enough to elevate damage to levels that can lead to cracking, even at denting depths of 10\% of the outer diameter.

\end{itemize}

Based on the data available and the (realistic) conditions considered here, these findings suggest
that external defects, such as dents and gouges, that are deemed safe in the operation of natural gas pipelines, will not constitute a structural integrity concern with the introduction of hydrogen. While this conclusion does not hold for a hydrogen gas pipeline at a high pressure (above 10 MPa), containing a pre-existing (passive) dent with a gouge, and where hydrogen egress is completely hindered, this $\textbf{j}^\mathrm{out}=0$ condition is very difficult to achieve in practice, particularly when that outer surface is undergoing or has undergone significant denting. However, it remains to assess the role that other factors, such as external cracks (e.g., originating from stress corrosion cracking) and fatigue damage, can play. Full-scale experimental validation of these findings is forthcoming.

\section*{Acknowledgments}
\label{Acknowledge of funding}

\noindent The authors acknowledge stimulating discussions with Sarah Hopkin (formerly at Shell) and data from the EU commission RFCS2027 project \#101112650: `Guidelines for material selection and qualification for safe transportation of H2-NG mixtures in EU pipelines'. The authors acknowledge financial support from Shell through the R\&D project ``Simulation-based assessment of the interplay between hydrogen, defects and plastic straining'' (\textit{SimHdefect}). E. Mart\'{\i}nez-Pa\~neda additionally acknowledges financial support from UKRI’s Future Leaders Fellowship programme [grant MR/V024124/1].

%% The Appendices part is started with the command \appendix;
%% appendix sections are then done as normal sections
%% If you have bibdatabase file and want bibtex to generate the
%% bibitems, please use
%%
%%  \bibliographystyle{elsarticle-harv} 
%%  \bibliography{<your bibdatabase>}

%\appendix

%% else use the following coding to input the bibitems directly in the
%% TeX file.

%\bibliographystyle{elsarticle-harv}
%\bibliography{biblio}

\begin{singlespace}
\raggedright

\end{singlespace}

\end{document}